\documentclass[namedreferences]{SolarPhysics}
\usepackage[optionalrh,solaenum]{spr-sola-addons} 
\usepackage{graphicx}                    
\usepackage{color}                       
\usepackage{url}                         

\usepackage[unicode=true,colorlinks=no,pdfborder=no]{hyperref}
\makeatletter
\AtBeginDocument{\Hy@breaklinkstrue}
\makeatother

\begin{document}

\begin{article}

\begin{opening}

\title{A Statistical Study on Photospheric Magnetic Nonpotentiality of Active Regions and Its Relationship with Flares during Solar Cycles 22--23}

%
\author{Xiao~\surname{YANG}$^{1,2}$\sep
        HongQi~\surname{ZHANG}$^{1}$\sep
        Yu~\surname{GAO}$^{1}$\sep
        Juan~\surname{GUO}$^{1}$\sep
        GangHua~\surname{LIN}$^{1}$
       }

%
\runningauthor{X.~\surname{YANG} {\it et al.}}
\runningtitle{Statistical Study on Magnetic Nonpotentiality and Associated Flares}

%
\institute{$^{1}$ Key Laboratory of Solar Activity, National Astronomical Observatories, Chinese Academy of Sciences, 20A Datun Road, Beijing 100012, P. R. China \\
           e-mail: \url{yangx@nao.cas.cn} \\
           $^{2}$ Graduate University of Chinese Academy of Sciences, 19A Yuquan Road, Beijing 100049, P. R. China
             }

\begin{abstract}\\
A statistical study is carried out on the photospheric magnetic nonpotentiality in solar active regions and its relationship with associated flares. We select 2173 photospheric vector magnetograms from 1106 active regions observed by the Solar Magnetic Field Telescope at Huairou Solar Observing Station, National Astronomical Observatories of China, in the period of 1988--2008, which covers most of the 22nd and 23rd solar cycles. We have computed the mean planar magnetic shear angle ($\overline{\Delta\phi}$), mean shear angle of the vector magnetic field ($\overline{\Delta\psi}$), mean absolute vertical current density ($\overline{|J_{z}|}$), mean absolute current helicity density ($\overline{|h_{\rm c}|}$), absolute twist parameter ($|\alpha_{\rm av}|$), mean free magnetic energy density ($\overline{\rho_{\rm free}}$), effective distance of the longitudinal magnetic field ($d_{\rm E}$), and modified effective distance ($d_{\rm Em}$) of each photospheric vector magnetogram. Parameters $\overline{|h_{\rm c}|}$, $\overline{\rho_{\rm free}}$, and $d_{\rm Em}$ show higher correlation with the evolution of the solar cycle. The Pearson linear correlation coefficients between these three parameters and the yearly mean sunspot number are all larger than 0.59. Parameters $\overline{\Delta\phi}$, $\overline{\Delta\psi}$, $\overline{|J_{z}|}$, $|\alpha_{\rm av}|$, and $d_{\rm E}$ show only weak correlations with the solar cycle, though the nonpotentiality and the complexity of active regions are greater in the activity maximum periods than in the minimum periods. All of the eight parameters show positive correlations with the flare productivity of active regions, and the combination of different nonpotentiality parameters may be effective in predicting the flaring probability of active regions.
\end{abstract}

%

\end{opening}

%

\section{Introduction}
     \label{S-introduction}

It is generally accepted that the magnetic field dominates the evolution of the Sun, and the magnetic nonpotentiality may be responsible for massive energy released in solar explosive activities such as flares and coronal mass ejections (CMEs). Even though the physical parameters that characterize the nonpotentiality of the magnetic fields do not directly trigger solar eruption, active regions (ARs) with strong nonpotentiality and considerable complexity are more likely to explode than simpler ones because the former have sufficient free energy to release. Studying the evolution of magnetic nonpotentiality and its relationship with solar eruption from a statistical point of view is useful in predicting eruptive events and in understanding the long-term evolution of solar activity.

The characteristics of nonpotential magnetic fields associated with flares in ARs have been studied for a couple of decades. Several nonpotential parameters of flare-productive ARs are investigated by many authors in detail. \inlinecite{Gary87} analyzed nonpotential features based on the observed vector magnetic field of NOAA AR 2684, and tried to answer the questions such as the relation between the nonpotential features and the flares. \inlinecite{WangJX96} examined in detail the relationship between flare occurrence and nonpotentiality development in a flare-productive region. \inlinecite{Moon00} analyzed the spatial distribution and temporal evolution of several nonpotential parameters in NOAA AR 5747. \inlinecite{Deng01} studied the daily evolution of magnetic nonpotentiality focusing on NOAA AR 9077 that produced a giant flare from 11 to 15 July 2000. \inlinecite{Leka03} investigated the magnitudes and temporal variations of several photospheric magnetic parameters, such as horizontal magnetic gradient, vertical current, current helicity, twist parameter $\alpha$, magnetic shear angles, and excess magnetic energy in three ARs, to identify properties that are important to the solar flare production. \inlinecite{Dun07} calculated the magnetic shear angle, vertical current, and current helicity of NOAA AR 10486 and analyzed their temporal evolutions and spatial relations to the flares. \citeauthor{Falconer02} (\citeyear{Falconer02}, \citeyear{Falconer06}) also discussed the relationship between several nonpotentiality-related parameters and CME productivity of ARs. Some authors studied the physical characteristics of photospheric magnetic fields in ARs and their relationship to flares in a statistical sense. \inlinecite{Cui06} preferred the maximum horizontal gradient, length of the neutral lines, and number of singular points; \inlinecite{Jing06} used the mean value of magnetic gradients at strong-gradient magnetic neutral lines, length of strong-gradient magnetic neutral lines, and total magnetic energy; \inlinecite{Cui07} also tried the length of neutral lines with strong gradient and/or strong shear; \inlinecite{Cui08} studied the total unsigned current and total unsigned current helicity. \citeauthor{Guo06} (\citeyear{Guo06}, \citeyear{Guo07a}, \citeyear{Guo10}) utilized the effective distance that could roughly quantify the magnetic complexity of ARs, to study the evolution of magnetic complexity in ARs, its relationship with erupting activity of ARs, and its evolution with the solar cycles, respectively. A close relationship between magnetic nonpotentiality of ARs and solar eruptions is observationally studied in numerous articles. \inlinecite{Rust94} reviewed the progress of pre-flare state studies, including nonpotential characteristics, and pointed out the importance of the more quantitative results from the vector magnetographs.

Lacking of long-term steady vector magnetic field observations, the line-of-sight magnetograms are mostly used as means of case analyses and statistical researches. It is necessary to study the long-term evolution of appropriate parameters derived from vector magnetic fields and find whether there is a correlation between these parameters and flares or other eruption events. So far ground-based telescopes, whose maintenance for long-term observations is relatively easy, still have the advantages in data accumulation. Since 1987, the Solar Magnetic Field Telescope (SMFT) \cite{Ai86}, installed at Huairou Solar Observing Station (HSOS), National Astronomical Observatories of China (NAOC), has reliably worked for more than twenty years that almost covers two complete solar cycles. Based on the long-term observations of SMFT, a statistical study on photospheric magnetic nonpotential parameters in ARs is carried out, and the relationship between these parameters and solar flares is also presented.

In Section \ref{S-observation}, observations by SMFT and data reductions are introduced. Measures of nonpotentiality and complexity are shown in Section \ref{S-param}. In Section \ref{S-analysis}, statistical analysis and results on magnetic nonpotentiality associated with flares are presented. Conclusions are given in Section \ref{S-conclusion}.

\section{Observations and Data Reductions}
     \label{S-observation}

\subsection{Observations at Huairou Solar Observing Station}
     \label{S-obsHSOS}
SMFT is a 35cm-aperture filter-type vector magnetograph that measures solar magnetic and velocity fields in both the photosphere and chromosphere. The magnetograms used in this study are obtained by one of its working spectral lines, Fe {\sc i} $\lambda$5324.19 \textrm{\AA} that measures photospheric vector magnetic fields. The Fe {\sc i} $\lambda$5324.19 \textrm{\AA} line is a strong and broad line with an equivalent width of about 0.334 \textrm{\AA} and a Land\'{e} factor of 1.5 \cite{Ai82,WangJX96}. Measuring at an offset $-$0.075 \textrm{\AA} from the line center of Fe {\sc i} $\lambda$5324.19 \textrm{\AA}, SMFT derives the longitudinal component of the vector magnetic field from the circular polarization (Stokes parameter $V$), and the transverse components from the linear polarization (Stokes parameters $Q$ and $U$) at the line center. It takes about 3 min to get the data for generating a vector magnetogram. The observations aim at limited areas of the solar disk, being mainly the ARs where sunspots are located. From 1988 to 2008, three different CCDs were used in SMFT, and the data were captured by them with three different sizes of the field of view (FOV): before 25 August 2001, the FOV was $5.23'\times3.63'$ and the pixel scale was $0.61''\times0.43''$; from 25 August 2001 to 30 November 2001, the FOV was $4.06'\times2.77'$ and the pixel size was $0.48''\times0.32''$; after 1 December 2001, the FOV was $3.75'\times2.81'$ with the pixel resolution of $0.35''\times0.35''$. To improve the signal-to-noise ratio (SNR), the observing system adopts a 256-frame integration for routine observations. After smoothing the three sets of data with $3\times5$, $3\times5$, and $5\times5$ pixels, respectively, we get a better SNR and the spatial resolution has been reduced to about $2''\times2''$. The $3\sigma$ noise of a vector magnetogram is about 20 gauss (G) for the longitudinal field and 150 \textrm{G} for the transverse field.

The vector magnetograms obtained from various magnetographs of different observatories were compared for the research of magnetic field structures in specific ARs \cite{WangHM92,Bao00,ZhangHQ03} and for the statistical analysis \cite{Pevtsov06,Xu07}, to confirm the accuracy of the magnetic field measurement by SMFT. As a result, these data are basically consistent, including the longitudinal and transverse components of the magnetic field, electric current, helicity parameters, and so on. Small discrepancies due to observing methods and data inversion methods are insignificant in the present statistical study of magnetic nonpotentiality.

\subsection{Data Selection and Preprocessing}
     \label{S-data}

The samples located within $30\,^{\circ}$ from the solar disk center have been selected to minimize the influence of projection effects. Of all the vector magnetograms, only one magnetogram is picked up for an AR each day. Most of selected samples are observed during 01:00 to 07:00 UT when HSOS is in a relatively better seeing condition. 2173 observed photospheric vector magnetograms containing 1106 ARs from June 1988 to March 2008 are chosen as samples. Table \ref{T-noMA} lists the selected magnetograms and ARs each year in the two cycles. Various types of ARs are included in this sample table, from simple unipolar regions to complex $\delta$ ARs. From 1988 to 2008, there are 6095 ARs recorded by NOAA, and the percentages of the ARs that produced $\geq$C-, $\geq$M-, and X-class flares are 39.1$\%$, 15.0$\%$, and 1.0$\%$, respectively. Our selected ARs account for 18.1$\%$ of the total ARs over this period, and 71.3$\%$, 20.7$\%$, and 1.0$\%$ of the selected ARs produced $\geq$C-, $\geq$M-, and X-class flares. Though these ARs are biased to the larger and more active regions, they only produced 18.6$\%$ of all the AR-related $\geq$C-class flares (C-class flares, 18.8$\%$; M-class flares, 17.2$\%$; X-class flares, 19.8$\%$) within the subsequent 48 h from the time of the selected magnetogram. Therefore, the sample coverage is reasonable and representative.

\begin{table}
\caption{Number distributions of selected magnetograms and active regions from 1988 to 2008. The counts of the magnetograms (MGs) and active regions (ARs) in the solar maximum periods are relatively higher than those in the solar minimum periods during the same solar cycle. The samples have reasonably covered two solar cycles.}
\label{T-noMA}
\tabcolsep=4.0pt
\begin{tabular}{cccccccccccc}
  \hline
    & 1988 & 1989 & 1990 & 1991 & 1992 & 1993 & 1994 & 1995 & 1996 & 1997 & 1998 \\
  \hline
MGs & 10 & 37 & 47 & 108 & 138 & 97 & 74 & 56 & 11 & 54 & 114 \\
ARs & 8 & 21 & 30 & 58 & 65 & 52 & 39 & 25 & 6 & 32 & 62 \\
  \hline
  \hline
    & 1999 & 2000 & 2001 & 2002 & 2003 & 2004 & 2005 & 2006 & 2007 & 2008 & Total \\
  \hline
MGs & 166 & 361 & 285 & 233 & 144 & 67 & 101 & 46 & 20 & 4 & 2173 \\
ARs & 88 & 166 & 136 & 123 & 77 & 34 & 50 & 22 & 11 & 2 & 1106 \\
  \hline
\end{tabular}
\end{table}

The one-to-one match between Huairou and NOAA AR numbers is manually done for analyzing the relationship between the magnetic characteristics of an AR and associated flares. Thus the flare records taken by {\it Geostationary Operational Environmental Satellites} (GOES) can be used in the study. As the FOV of SMFT is fixed, each magnetogram containing a single complete AR near the center of the FOV is sampled. Therefore, excluding the possible appearance of other ARs in a particular magnetogram, the data reduction is uniformly carried out on all the selected magnetograms. The lists of soft X-ray (SXR) flare events are recorded by GOES in detail, which can be downloaded from National Geophysical Data Center (NGDC)\footnote{\url{ftp://ftp.ngdc.noaa.gov/STP/SOLAR_DATA/SOLAR_FLARES/XRAY_FLARES/}}.

A vector magnetogram consists of the longitudinal ($B_{\ell}$) and transverse ($B_{\rm t}$) components of the magnetic field, after processing the raw data of Stokes $I$, $Q$, $U$, and $V$ signals. The calibration method commonly used for SMFT is based on the relationship
\begin{eqnarray}
B_{\ell}=C_{\ell}\frac{V}{I},\ \ \ \ \ \ B_{\rm t}=C_{\rm t}\sqrt[4]{\left(\frac{Q}{I}\right)^{2}+\left(\frac{U}{I}\right)^{2}},
\end{eqnarray}
where $I$, $Q$, $U$, and $V$ are Stokes parameters, and $C_{\ell}$ and $C_{\rm t}$ are the corresponding calibration coefficients of $B_{\ell}$ and $B_{\rm t}$. The azimuth angle ($\phi$) of the transverse field is obtained by
\begin{eqnarray}
\phi=\frac{1}{2} \arctan\left(\frac{U}{Q}\right).
\end{eqnarray}
From several calibrations on SMFT vector magnetographs \cite{Ai82,WangTJ96,Su04}, $C_{\ell}=8381$ and $C_{\rm t}=6790$ \cite{Su04} are selected in the present study. The statistical results are not affected significantly by the coefficient selection.

The potential field approximation method \cite{Harvey69,Sakurai85} is applied to resolve the $180\,^{\circ}$ ambiguity of the transverse field. This method works well for the regions where the actual magnetic shear angles are less than $90\,^{\circ}$. Though the azimuth ambiguity resolution would be rough in a few regions that are very nonpotential, the analyzed parameters we use are all macroscopic and averaged quantities over a whole AR, thus the influence could be neglected in the statistical analysis. All the magnetograms are not deprojected because the selected samples are located near the solar disk center \cite{Gary90,LiH02,Cui07}.

\section{Magnetic Nonpotentiality and Complexity Parameters}
     \label{S-param}

\subsection{Magnetic Shear Angle}
     \label{S-shearangle}
The magnetic shear is one of the typical parameters to describe the nonpotentiality of magnetic fields in the solar photosphere. \inlinecite{Hagyard84} and \inlinecite{Lv93} introduced the planar and three-dimensional magnetic shear angles, respectively. The three-dimensional magnetic shear angle $\Delta\psi$ is the angle between the directions of the observed vector magnetic field $\emph{\textbf{B}}_{\rm o}$ and its corresponding potential field $\emph{\textbf{B}}_{\rm p}$. The planar magnetic shear $\Delta\phi$, which can be considered as the projection of $\Delta\psi$ onto the plane of the sky, is defined as the azimuthal difference between the observed field and the potential field. $\Delta\psi$ is a more direct measure of deviation from a potential field than $\Delta\phi$, and has a direct relation to the magnetic free energy of the nonpotential fields as well (see Section \ref{S-fed}; \opencite{Lv93}). Here $\Delta\phi$ and $\Delta\psi$, which are the angles between the directions of two vectors, are both without sign:
\begin{eqnarray}
\Delta\phi=(\widehat{\emph{\textbf{B}}_{\rm to},\emph{\textbf{B}}_{\rm tp}})=\arccos\left(\frac{\emph{\textbf{B}}_{\rm to}\cdot\emph{\textbf{B}}_{\rm tp}}{|\emph{\textbf{B}}_{\rm to}||\emph{\textbf{B}}_{\rm tp}|}\right),
\label{Eq-dPhi}
\end{eqnarray}
\begin{eqnarray}
\Delta\psi=(\widehat{\emph{\textbf{B}}_{\rm o},\emph{\textbf{B}}_{\rm p}})=\arccos\left(\frac{\emph{\textbf{B}}_{\rm o}\cdot\emph{\textbf{B}}_{\rm p}}{|\emph{\textbf{B}}_{\rm o}||\emph{\textbf{B}}_{\rm p}|}\right),
\label{Eq-dPsi}
\end{eqnarray}
where the subscripts o and p indicate the observed and potential magnetic fields, respectively, and t refers to the transverse component of the magnetic field.

\subsection{Vertical Current Density, Current Helicity Density, and Twist Parameter $\alpha$}

The electric current, current helicity, and twist parameter (force-free field factor) $\alpha$ are also important to describe magnetic nonpotentiality. Early in 1960s, solar physicists began to observationally study the electric currents in ARs \cite{Severny65} and their association with solar flares \cite{Moreton68}. \inlinecite{WangTJ94} and \inlinecite{Schrijver07} indicated that the emergence of magnetic fields carrying electric currents may cause most major solar flares. \inlinecite{Bao99} testified that the current helicity ({\it cf.}, \opencite{Seehafer90}; \opencite{Pevtsov95}) of ARs plays an important role in solar flare evolution. \inlinecite{Nakagawa72} theoretically described the force-free field factor $\alpha$ as an indicator of the twist of the magnetic field lines; they deduced that the magnetic energy increases with the value of $\alpha$. \inlinecite{Nindos04} discussed CME-associated big flares of 133 events to explain the roles of the magnetic helicity and force-free factor $\alpha$ of the pre-flare corona.

According to Ampere's law and neglecting the effect of the electric displacement current, the vertical component of electric current density is obtained by
\begin{eqnarray}
J_{z}=\frac{1}{\mu_{0}}(\nabla\times\emph{\textbf{B}})_{z}=\frac{1}{\mu_{0}}\left(\frac{\partial{B_{y}}}{\partial{x}}-\frac{\partial{B_{x}}}{\partial{y}}\right),
\label{Eq-Jz}
\end{eqnarray}
where $J_{z}$ is in units of \textrm{A km$^{-2}$}, $\mu_{0}=4\pi\times10^{-6}$ \textrm{G km A$^{-1}$} is the permeability in free space, and $B_{x}$ and $B_{y}$ are the two perpendicular components of a horizontal magnetic field.

The current helicity is defined as
\begin{eqnarray}
H_{\rm c}=\int_{V}\emph{\textbf{B}}\cdot(\nabla\times\emph{\textbf{B}})\ \mathrm{d}V,   \nonumber
\end{eqnarray}
and the vertical current helicity density can then be calculated by
\begin{eqnarray}
(h_{\rm c})_{z}=B_{z}(\nabla\times\emph{\textbf{B}})_{z}=B_{z}\left(\frac{\partial{B_{y}}}{\partial{x}}-\frac{\partial{B_{x}}}{\partial{y}}\right).
\label{Eq-hc}
\end{eqnarray}
So far there is no observational method to obtain the transverse component of current helicity density $(h_{\rm c})_{\rm t}$, and the symbol $h_{\rm c}$ is usually used to represent its vertical component.

On the basis of the force-free field assumption, the Lorentz force is equal to zero, {\it i.e.} the electric current is parallel to the local magnetic field:
\begin{eqnarray}
(\nabla\times\emph{\textbf{B}})\times\emph{\textbf{B}}=0,   \nonumber
\end{eqnarray}
or
\begin{eqnarray}
\nabla\times\emph{\textbf{B}}=\alpha\emph{\textbf{B}}.   \nonumber
\end{eqnarray}
If the magnetic field is approximated to be a linear force-free field, the local twist $\alpha$ is expressed as
\begin{eqnarray}
\alpha=\frac{(\nabla\times\emph{\textbf{B}})_{z}}{B_{z}}.   \nonumber
\end{eqnarray}
We use $\alpha_{\rm av}$ \cite{Hagino04} to characterize the dominated twist status of an AR:
\begin{eqnarray}
\alpha_{\rm av}=\frac{\sum(\nabla\times\emph{\textbf{B}})_{z}\cdot {\rm sign}[B_{z}]}{\sum|B_{z}|}.
\label{Eq-alpha}
\end{eqnarray}

\subsection{Free Magnetic Energy Density}
     \label{S-fed}
The energy released through solar flares and other explosive events relies on the accumulation of the free magnetic energy (nonpotential magnetic energy) that is defined as the difference between the total magnetic energy ($E$) and potential magnetic energy ($E_{\rm p}$):
\begin{eqnarray}
\Delta E = E-E_{\rm p}.   \nonumber
\end{eqnarray}
\inlinecite{Hagyard81} introduced the source field to describe the nonpotentiality of a magnetic field on the photosphere:
\begin{eqnarray}
\emph{\textbf{B}}_{\rm s}=\emph{\textbf{B}}_{\rm o}-\emph{\textbf{B}}_{\rm p},   \nonumber
\end{eqnarray}
where $\emph{\textbf{B}}_{\rm o}$ is the observed vector magnetic field, $\emph{\textbf{B}}_{\rm p}$ denotes the potential field extrapolated from the longitudinal component of $\emph{\textbf{B}}_{\rm o}$, and $\emph{\textbf{B}}_{\rm s}$ is the so-called source field which is the nonpotential component of the magnetic field.

The magnetic energy density of the source field is proportional to $\emph{\textbf{B}}_{\rm s}^{2}$:
\begin{eqnarray}
\rho_{\rm free}=\frac{\emph{\textbf{B}}_{\rm s}^{2}}{8\pi}=\frac{(\emph{\textbf{B}}_{\rm o}-\emph{\textbf{B}}_{\rm p})^{2}}{8\pi}.    \label{Eq-fED}
\end{eqnarray}
Equation~(\ref{Eq-fED}) is used to calculate the free magnetic energy density on the photospheric layer in the present work. It is deduced from Equation~(\ref{Eq-fED}) that
\begin{eqnarray}
\rho_{\rm free}=\frac{(B_{\rm o}-B_{\rm p})^{2}}{8\pi}+\frac{B_{\rm o}B_{\rm p}}{2\pi}\sin^{2} \left(\frac{\Delta\psi}{2}\right),   \nonumber
\end{eqnarray}
in which $B_{\rm o}=|\emph{\textbf{B}}_{\rm o}|$ and $B_{\rm p}=|\emph{\textbf{B}}_{\rm p}|$. We can then find that the shear angle of the vector magnetic field $\Delta\psi$ is directly related to the nonpotential magnetic energy \cite{Lv93,WangJX96}.

\subsection{Effective Distance}
Effective distance ($d_{\rm E}$), a structural parameter of an AR, proposed by \inlinecite{Chumak87}, presents a distinction between flare-quiet and flare-imminent ARs \cite{Chumak04}. As a quantified magnetic complexity, $d_{\rm E}$ depicts the degree of the isolation or mutual penetration of the two polarities of an AR in a geometrical sense \cite{Guo06}. $d_{\rm E}$ is calculated from
\begin{eqnarray}
d_{\rm E}=\frac{R_{\rm p}+R_{\rm n}}{R_{\rm pn}},
\label{Eq-dE}
\end{eqnarray}
where
\begin{eqnarray}
R_{\rm p}=\sqrt{\frac{A_{\rm p}}{\pi}},\ \ \ \ \ \ R_{\rm n}=\sqrt{\frac{A_{\rm n}}{\pi}}.   \nonumber
\end{eqnarray}
$A_{\rm p}$ ($A_{\rm n}$) is the total area of positive (negative) polarity, $R_{\rm p}$ ($R_{\rm n}$) is the equivalent radius of positive (negative) polarity, respectively, and $R_{\rm pn}$ is the distance between the flux-weighted centers of the two opposite polarities (\opencite{Guo06}, \citeyear{Guo07a}, \citeyear{Guo10}; \opencite{Guo07b}).

Considering the limitation of the parameter $d_{\rm E}$ to uniquely characterize the complexity of an AR, an additional factor $\overline{|B_{z}|}$ is multiplied to $d_{\rm E}$. The modified effective distance $d_{\rm Em}$ is defined as
\begin{eqnarray}
d_{\rm Em}=d_{\rm E} \overline{|B_{z}|}=\frac{R_{\rm p}+R_{\rm n}}{R_{\rm pn}} \overline{|B_{z}|}.
\label{Eq-dEm}
\end{eqnarray}
$d_{\rm E}$ is dimensionless, but $\overline{|B_{z}|}$ provides $d_{\rm Em}$ some practical physical meaning to some extent. Here, the dimension of $d_{\rm Em}$ is the same as that of the magnetic field, in the unit of gauss. $d_{\rm Em}$ also reflects the complexity of an AR, and could be considered as the weighted complexity of entire AR in terms of the averaged strength of the sunspot magnetic field.

\section{Statistical Analysis and Results}
     \label{S-analysis}

\subsection{Strength Distribution of Nonpotentiality during Solar Cycles 22--23}
     \label{S-npdistri}

\begin{figure}
  \centerline{\includegraphics[width=0.618\textwidth,clip=]{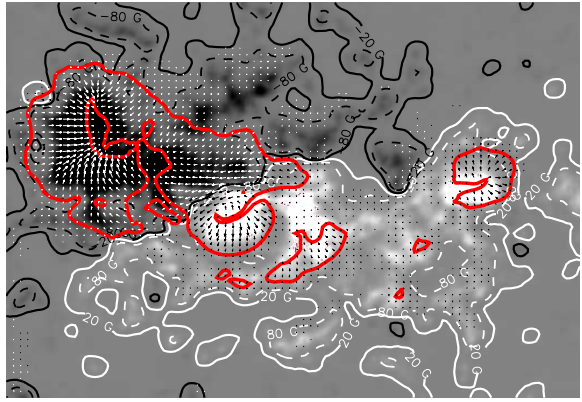}
             }
\caption{A vector magnetogram of NOAA AR 5356 at 02:12 UT on 15 February 1989. The white and black regions correspond to the positive and negative longitudinal magnetic fields, respectively. The arrows indicate the directions of transverse fields, the lengths of which denote their magnitudes. Only the arrow vectors with $B_{\rm t}>50$ \textrm{G} are shown in the figure. The white and black solid contours indicate $B_{\rm l}=\pm\,20$ \textrm{G}; the white and black dashed contours indicate $B_{\rm l}=\pm\,80$ \textrm{G}. The red solid contours are for $B_{\rm t}=200$ \textrm{G}. The FOV is about $5.23'\times3.63'$.}
     \label{F-maggram}
\end{figure}

By means of the magnetic nonpotentiality and complexity parameters described in Section \ref{S-param}, Equations (\ref{Eq-dPhi})--(\ref{Eq-dEm}) are used to calculate, for each photospheric vector magnetogram, the mean values of the planar magnetic shear angle ($\overline{\Delta\phi}$), shear angle of the vector magnetic field ($\overline{\Delta\psi}$), absolute vertical current density ($\overline{|J_{z}|}$), absolute current helicity density ($\overline{|h_{\rm c}|}$), and free magnetic energy density ($\overline{\rho_{\rm free}}$), and also calculate the absolute twist parameter ($|\alpha_{\rm av}|$), effective distance of the longitudinal magnetic field ($d_{\rm E}$), and modified effective distance ($d_{\rm Em}$). In each of the magnetograms, only the areas where the strength of longitudinal magnetic field is greater than 20 \textrm{G} are used in the calculations for the parameters $\overline{|J_{z}|}$, $\overline{|h_{\rm c}|}$, $|\alpha_{\rm av}|$, $\overline{\rho_{\rm free}}$, and $d_{\rm Em}$. The two mean shear angles $\overline{\Delta\phi}$ and $\overline{\Delta\psi}$ are obtained on the pixels with both the transverse magnetic field greater than 200 \textrm{G} and the strength of longitudinal field greater than 20 \textrm{G}, which are distributed at the sunspot penumbra and along the polarity inversion lines. The longitudinal field strength of 80 \textrm{G} is chosen as the lower threshold for obtaining $d_{\rm E}$. Figure \ref{F-maggram} shows one of the magnetogram samples, which is NOAA AR 5356 observed at 02:12 UT on 15 February 1989 by SMFT. The white and black solid contours ($B_{z}=\pm\,20$ \textrm{G}) depict the boundaries for calculating $\overline{\Delta\phi}$, $\overline{\Delta\psi}$, $\overline{|J_{z}|}$, $\overline{|h_{\rm c}|}$, $|\alpha_{\rm av}|$, $\overline{\rho_{\rm free}}$, and $d_{\rm Em}$. The white and black dashed contours ($B_{z}=\pm\,80$ \textrm{G}) enclose the regions for calculating $d_{\rm E}$. The red solid contours ($B_{\rm t}=200$ \textrm{G}) mark the edges for further confining the pixels used to obtain $\overline{\Delta\phi}$ and $\overline{\Delta\psi}$.

\begin{figure}
  \centerline{\hspace*{0.0\textwidth}
              \includegraphics[width=0.525\textwidth,clip=]{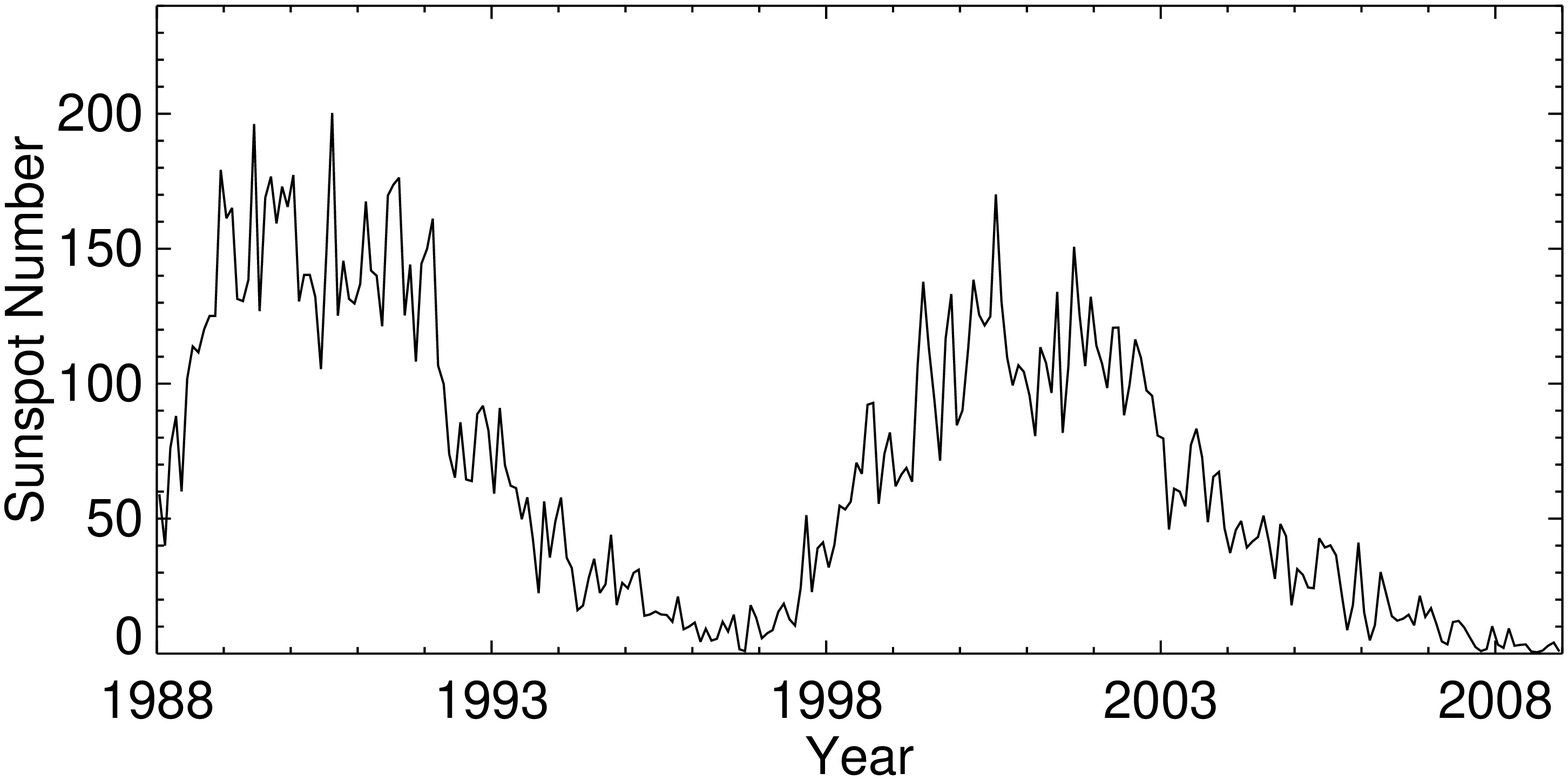}
              \hspace*{-0.025\textwidth}
              \includegraphics[width=0.525\textwidth,clip=]{ssn_bw.eps}
             }
  \centerline{\hspace*{0.0\textwidth}
              \includegraphics[width=0.525\textwidth,clip=]{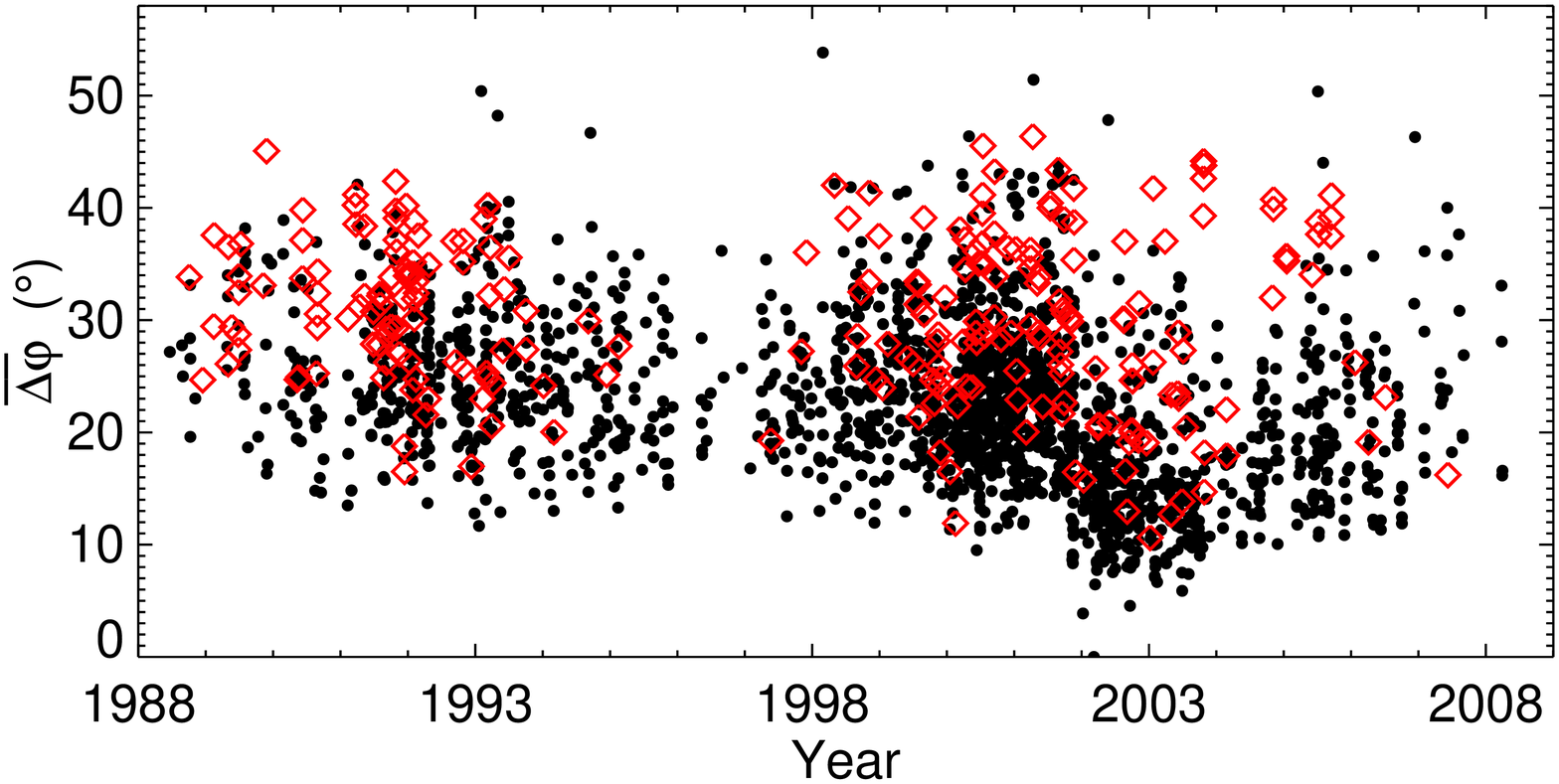}
              \hspace*{-0.025\textwidth}
              \includegraphics[width=0.525\textwidth,clip=]{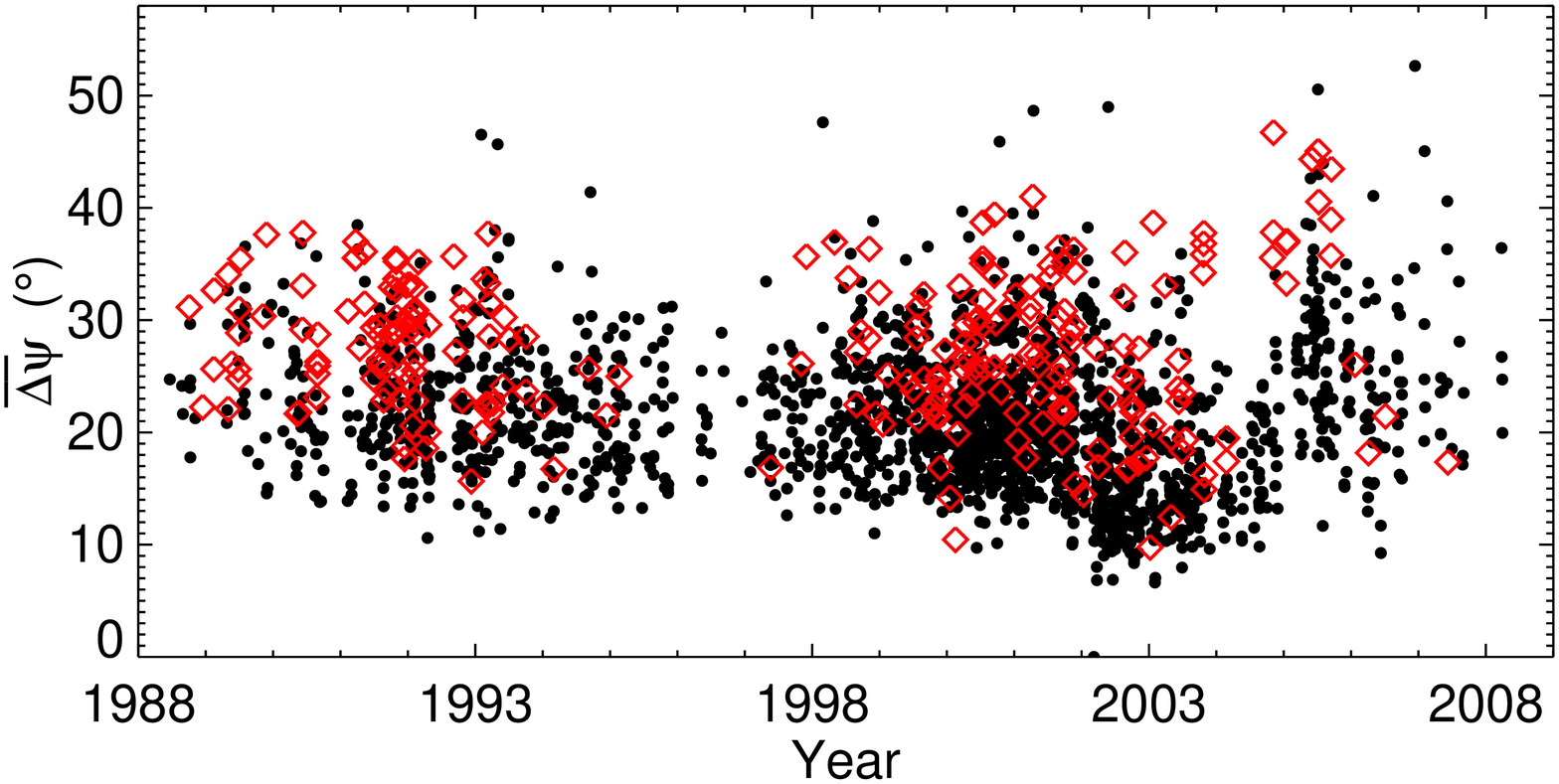}
             }
  \centerline{\hspace*{0.0\textwidth}
              \includegraphics[width=0.525\textwidth,clip=]{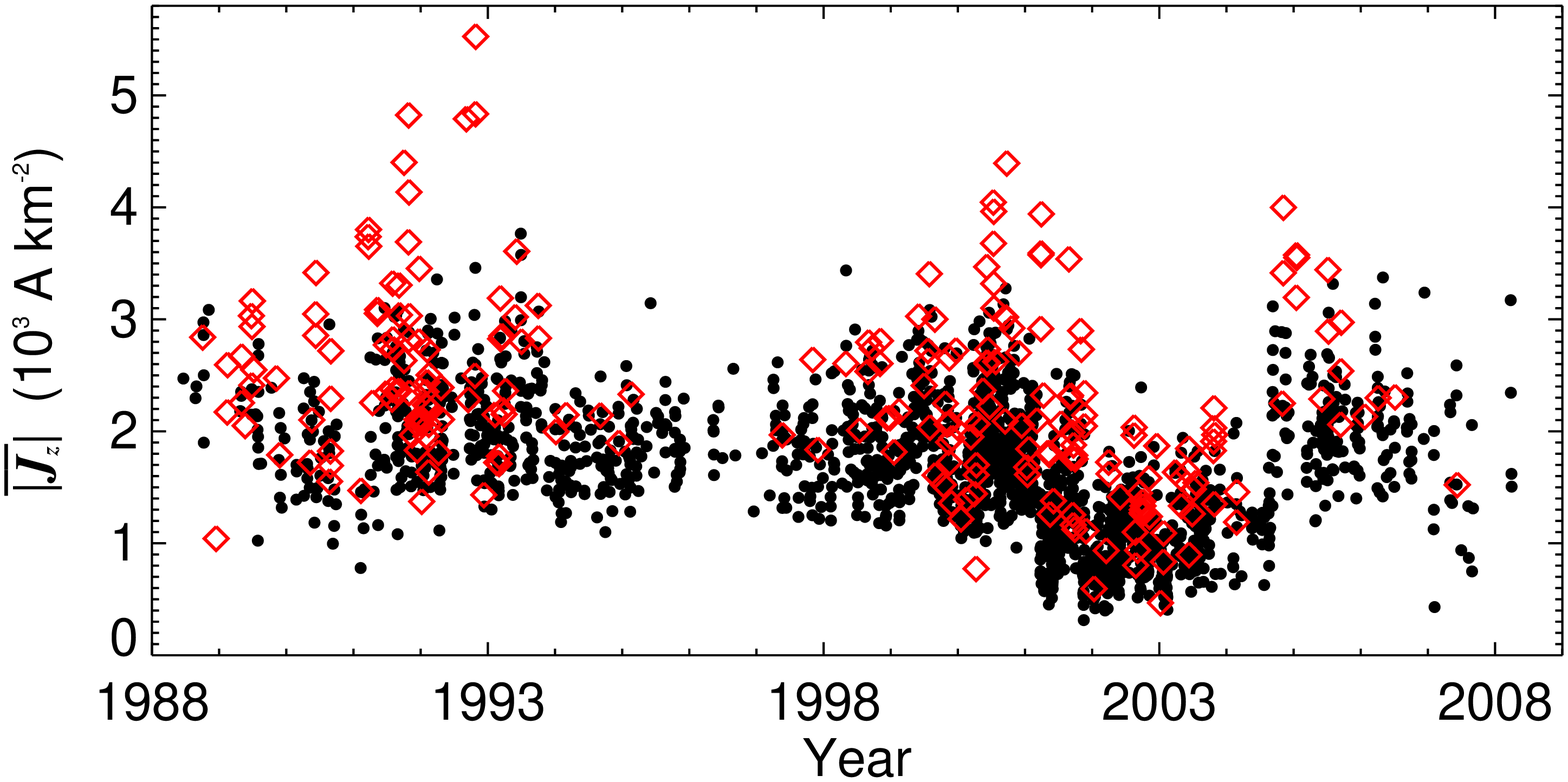}
              \hspace*{-0.025\textwidth}
              \includegraphics[width=0.525\textwidth,clip=]{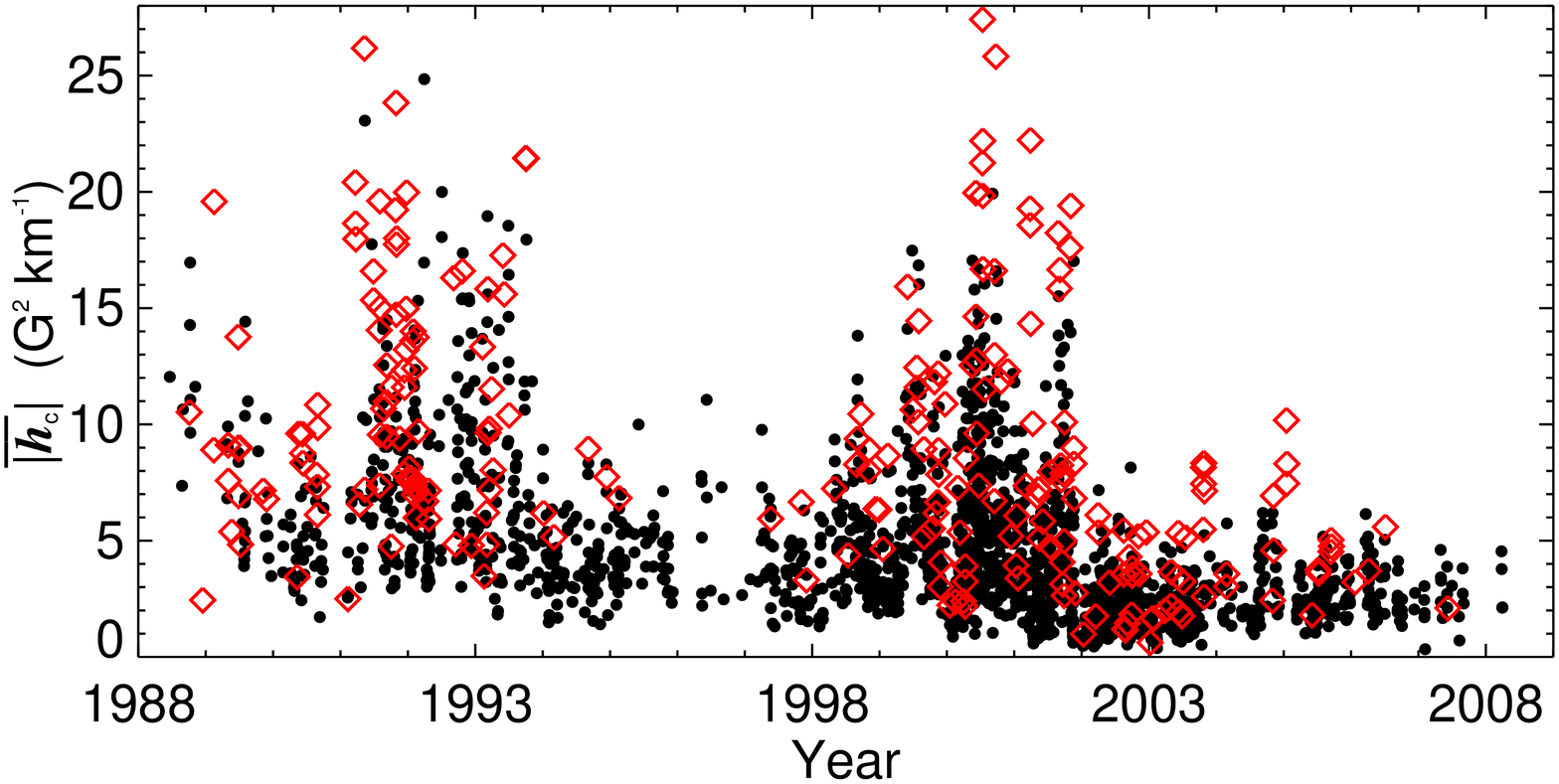}
             }
  \centerline{\hspace*{0.0\textwidth}
              \includegraphics[width=0.525\textwidth,clip=]{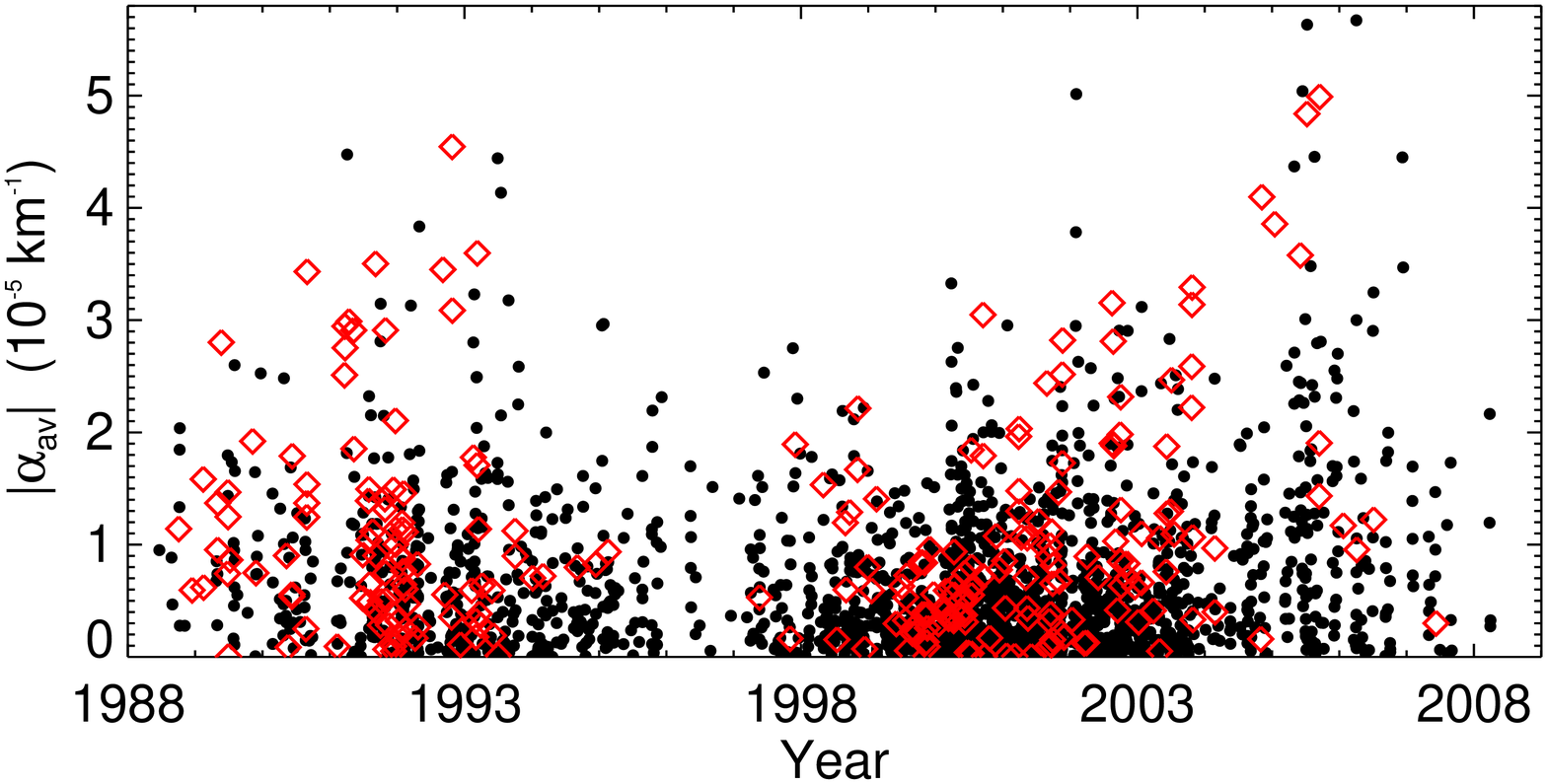}
              \hspace*{-0.025\textwidth}
              \includegraphics[width=0.525\textwidth,clip=]{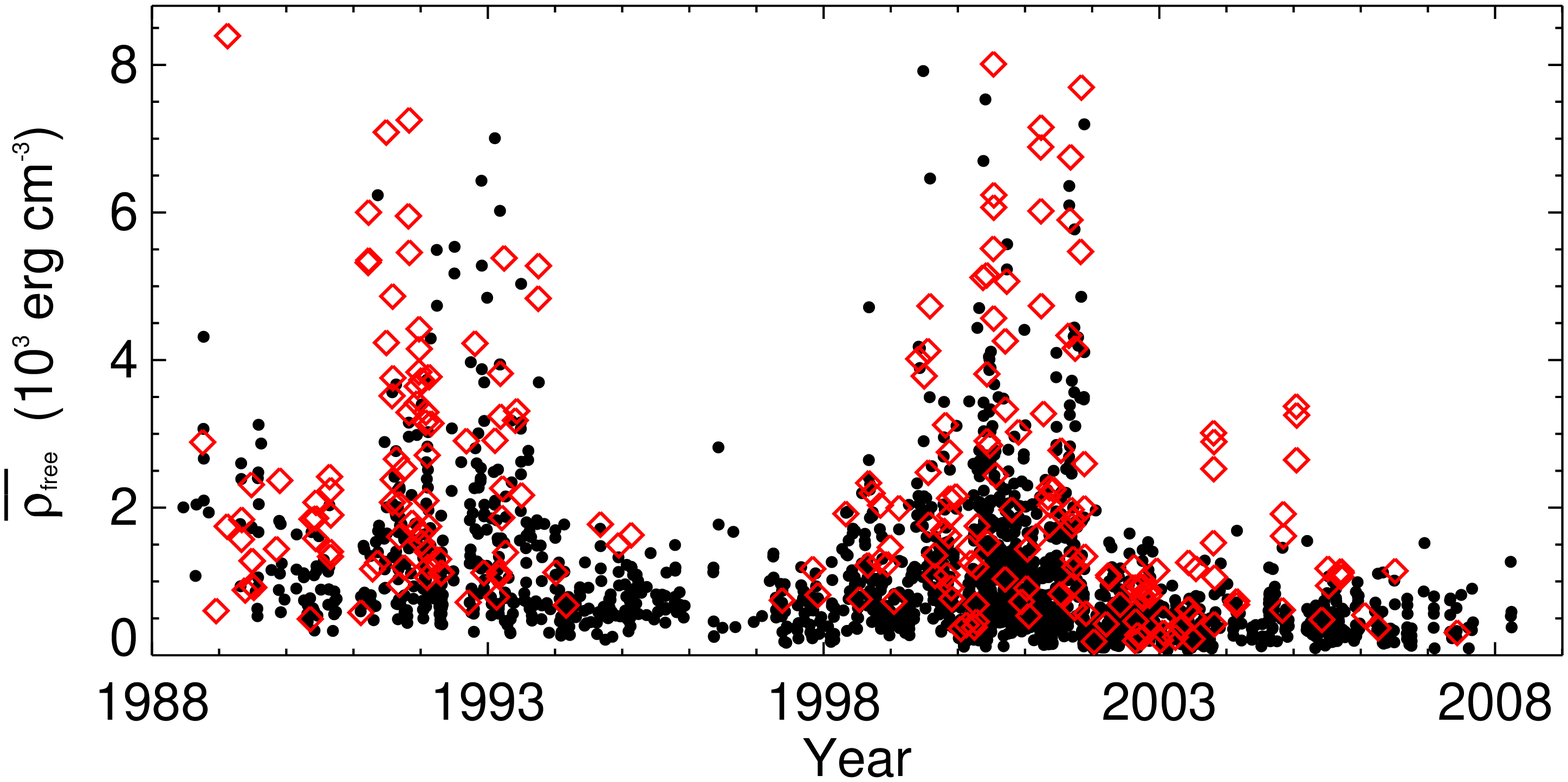}
             }
  \centerline{\hspace*{0.0\textwidth}
              \includegraphics[width=0.525\textwidth,clip=]{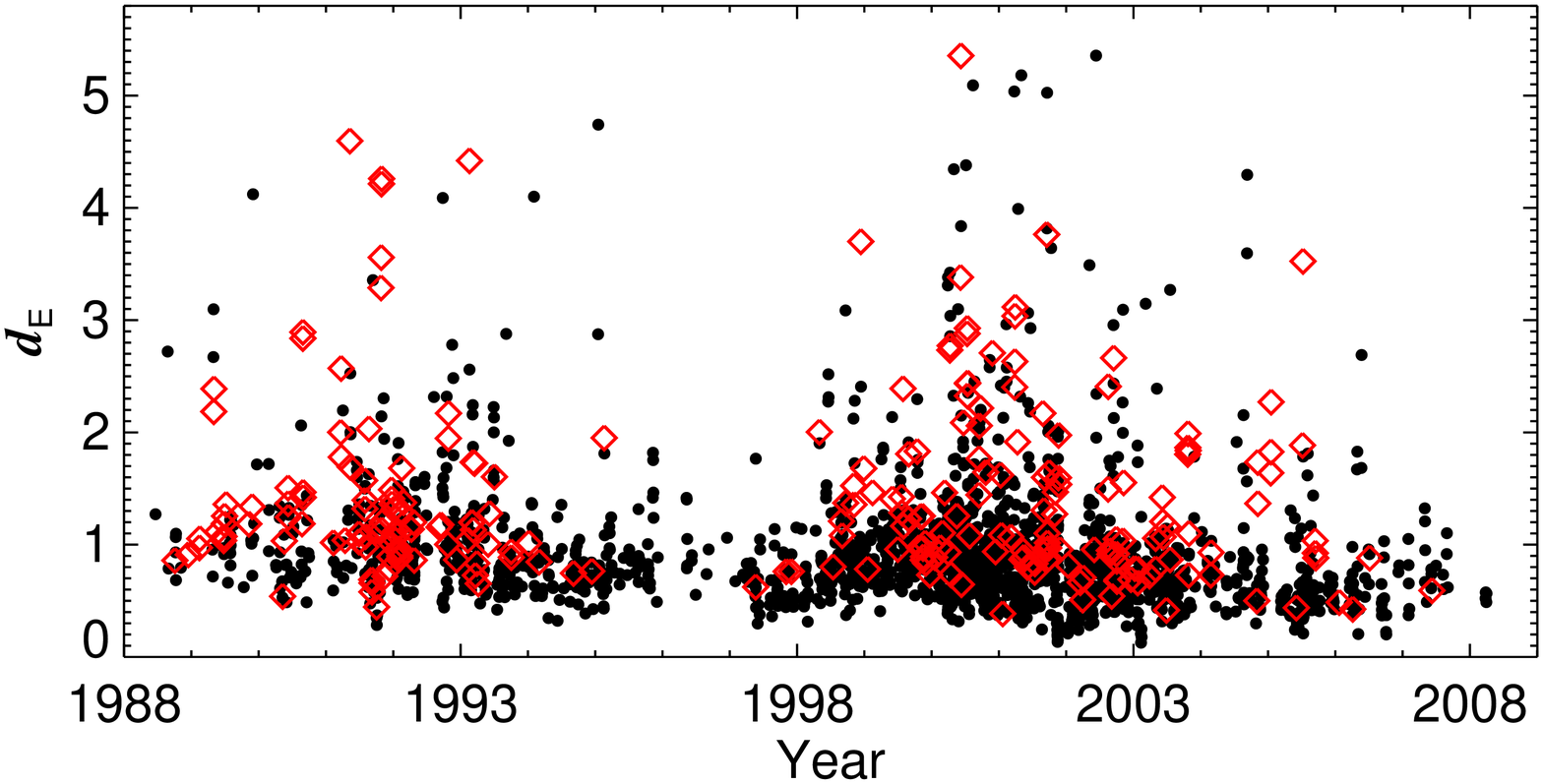}
              \hspace*{-0.025\textwidth}
              \includegraphics[width=0.525\textwidth,clip=]{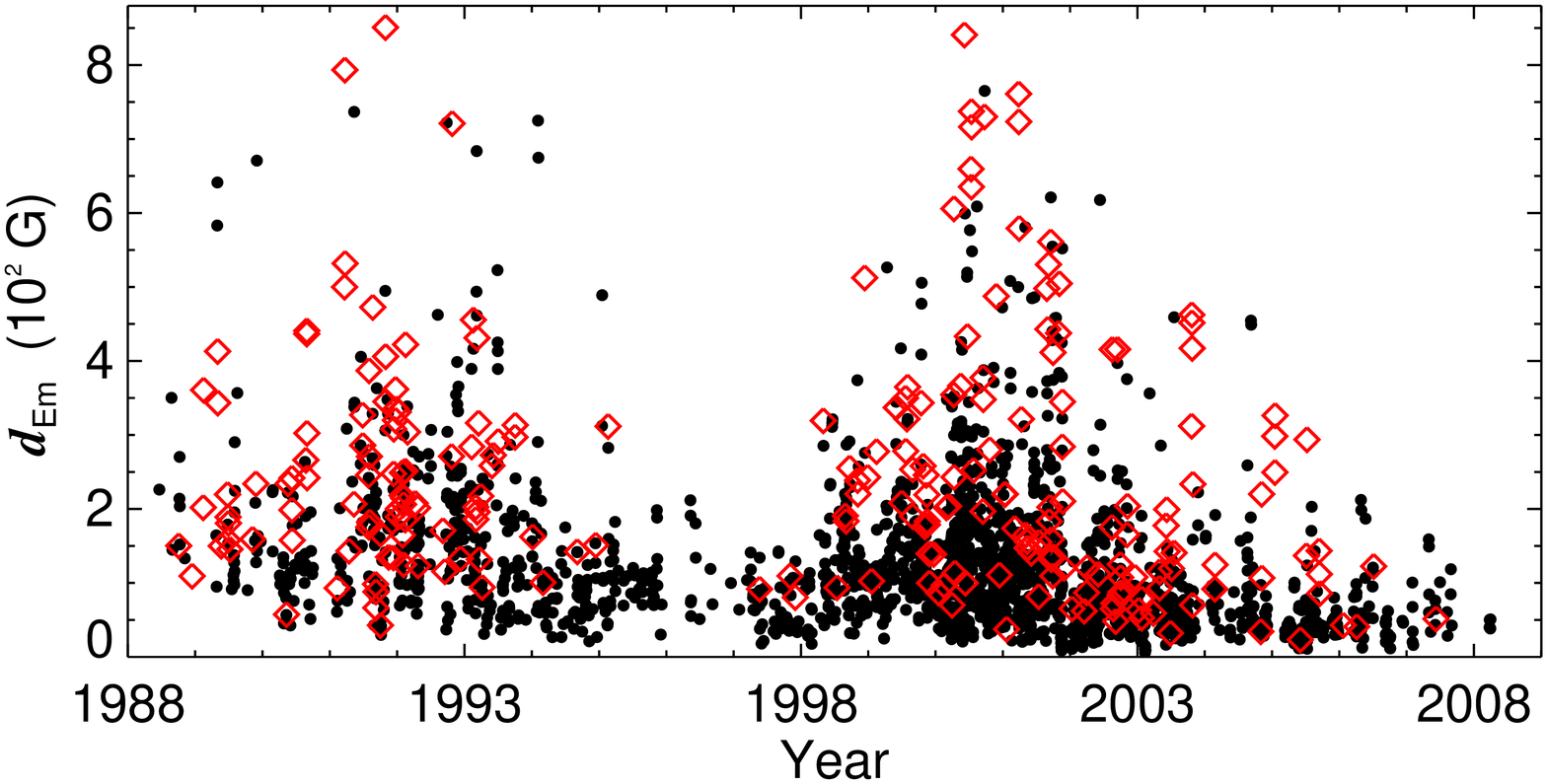}
             }
\caption{$\overline{\Delta\phi}$, $\overline{\Delta\psi}$, $\overline{|J_{z}|}$, $\overline{|h_{\rm c}|}$, $|\alpha_{\rm av}|$, $\overline{\rho_{\rm free}}$, $d_{\rm E}$, and $d_{\rm Em}$ of each AR sample during 1988--2008. Black dots represent the samples that did not produce flares with ${\rm FI} \geq 10.0$ in the following 24 h (flare-quiet samples). Red diamonds denote the samples that produced flares with ${\rm FI} \geq 10.0$ in the following 24 h (flare-productive samples). Two plots of monthly mean sunspot numbers during the same period are placed at the top of other panels for reference.}
     \label{F-scatpoint}
\end{figure}

\begin{figure}
  \centerline{\hspace*{0.0\textwidth}
              \includegraphics[width=0.525\textwidth,clip=]{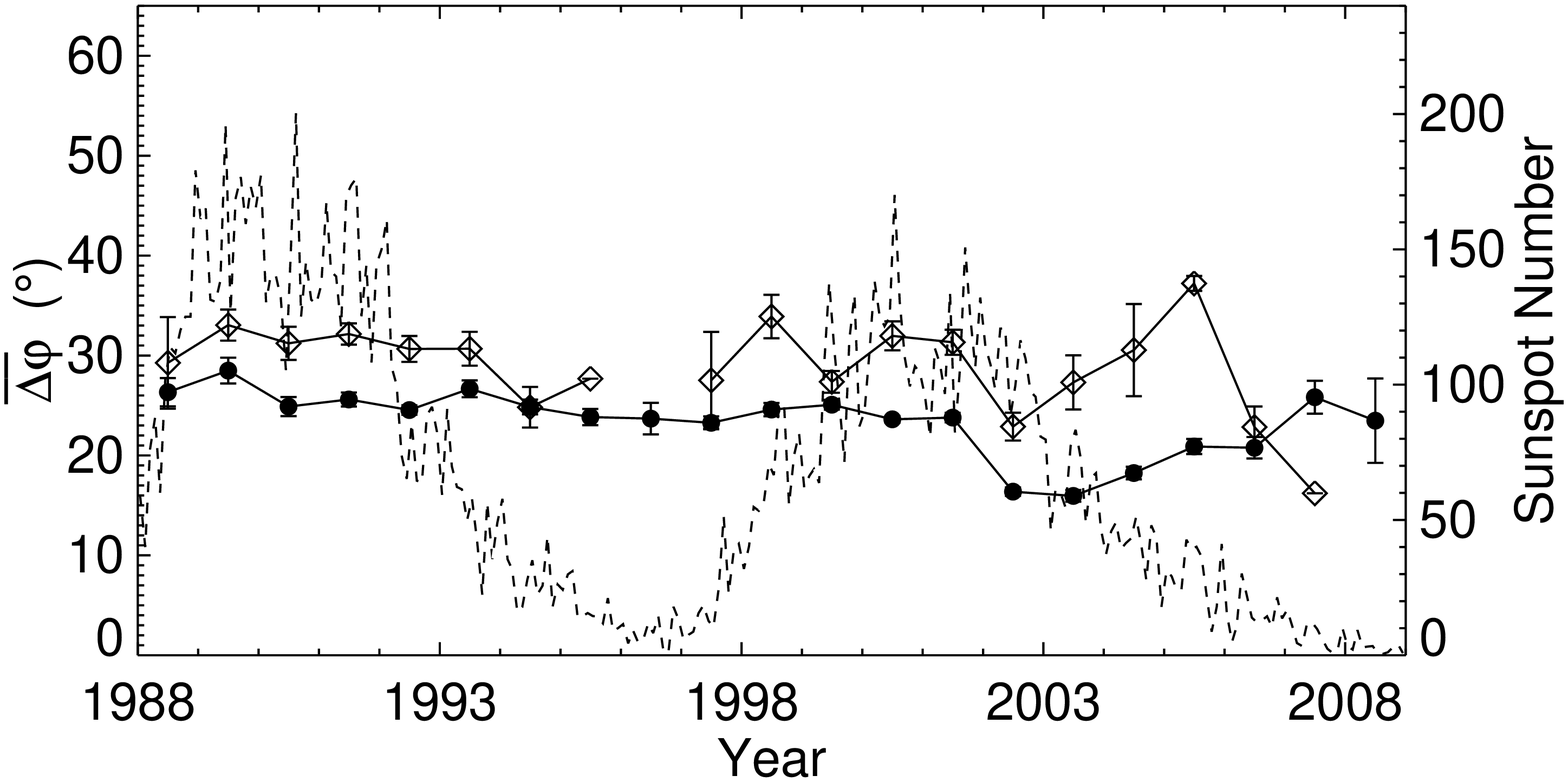}
              \hspace*{-0.025\textwidth}
              \includegraphics[width=0.525\textwidth,clip=]{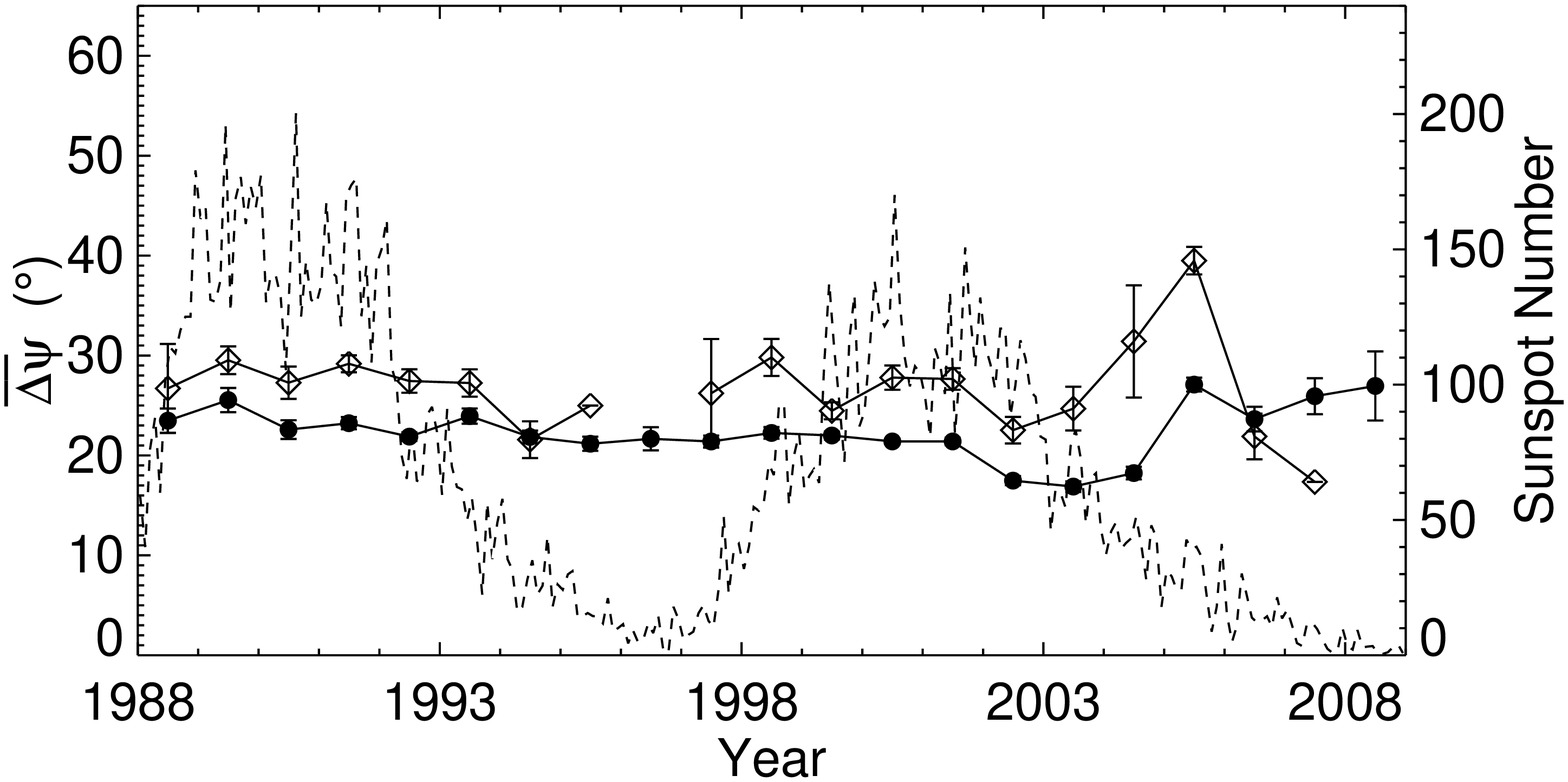}
             }
  \centerline{\hspace*{0.0\textwidth}
              \includegraphics[width=0.525\textwidth,clip=]{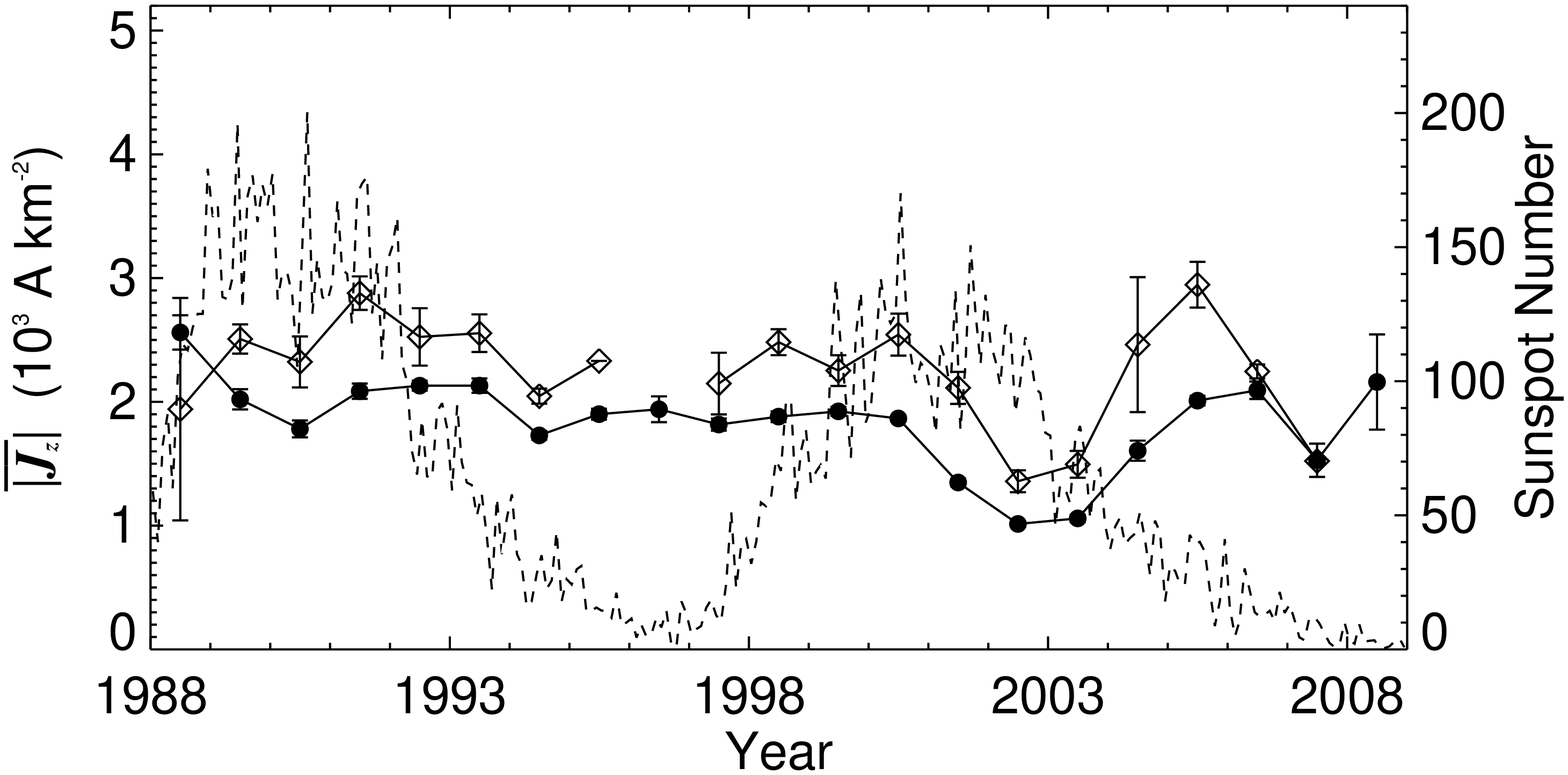}
              \hspace*{-0.025\textwidth}
              \includegraphics[width=0.525\textwidth,clip=]{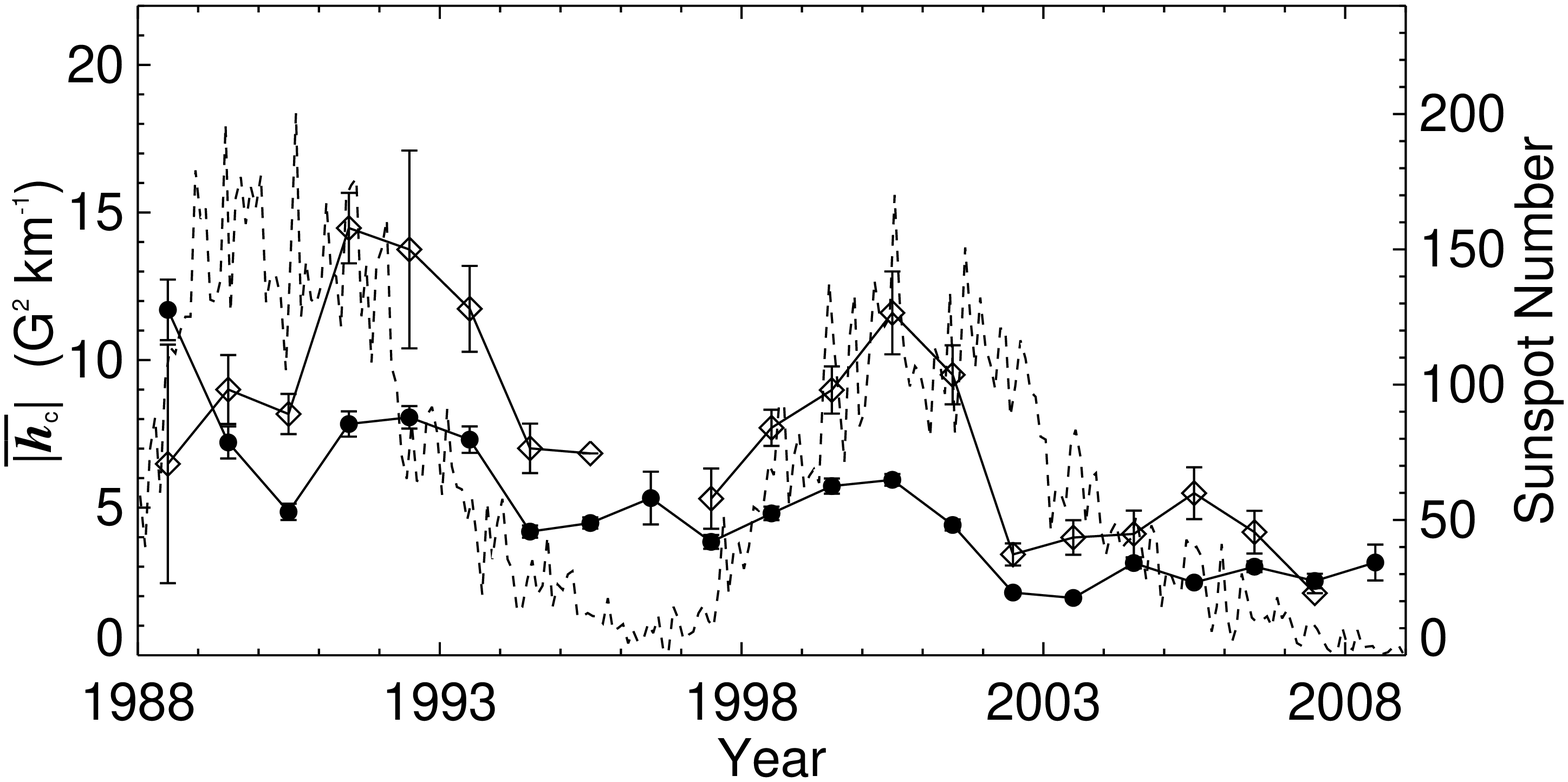}
             }
  \centerline{\hspace*{0.0\textwidth}
              \includegraphics[width=0.525\textwidth,clip=]{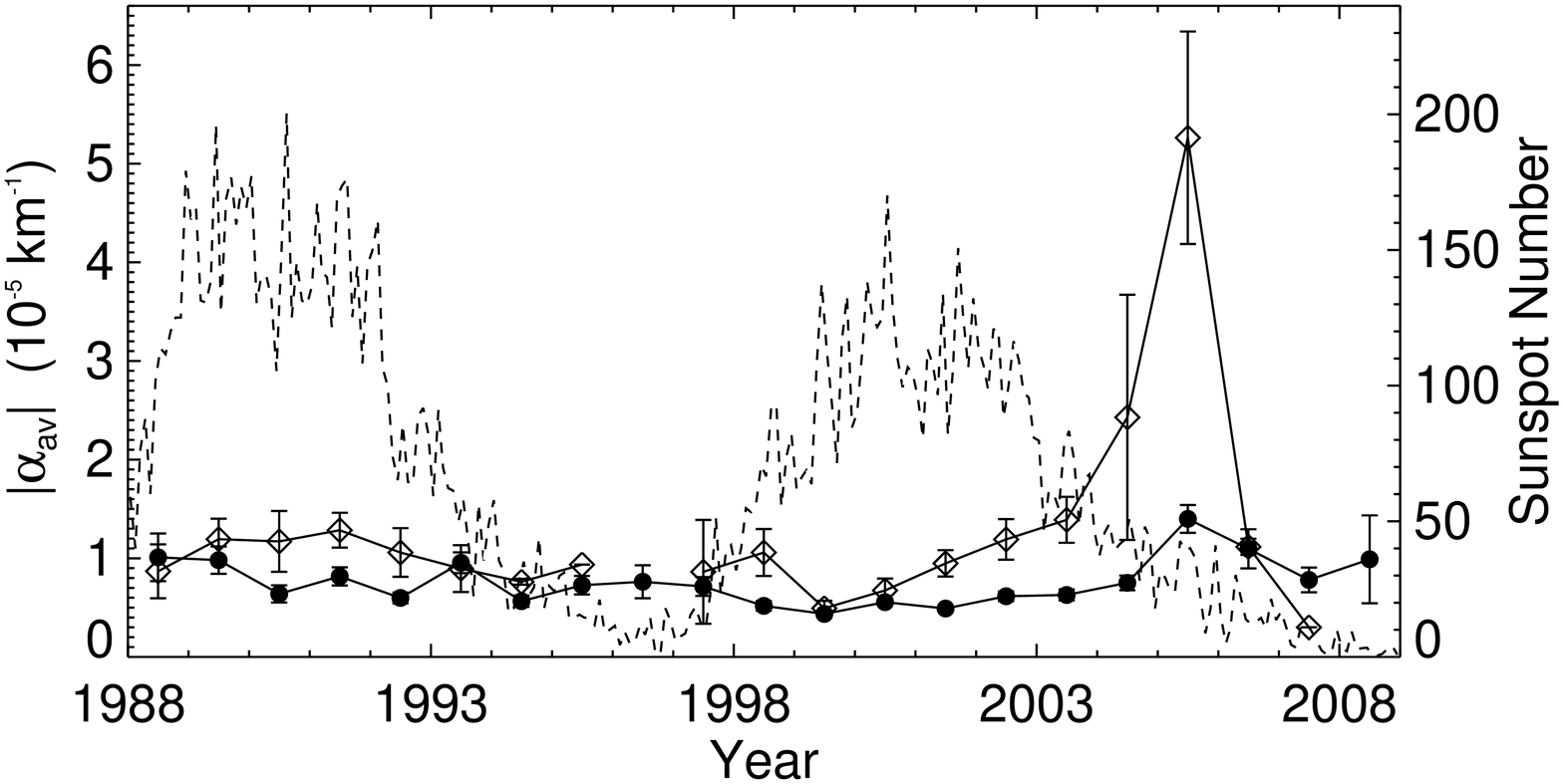}
              \hspace*{-0.025\textwidth}
              \includegraphics[width=0.525\textwidth,clip=]{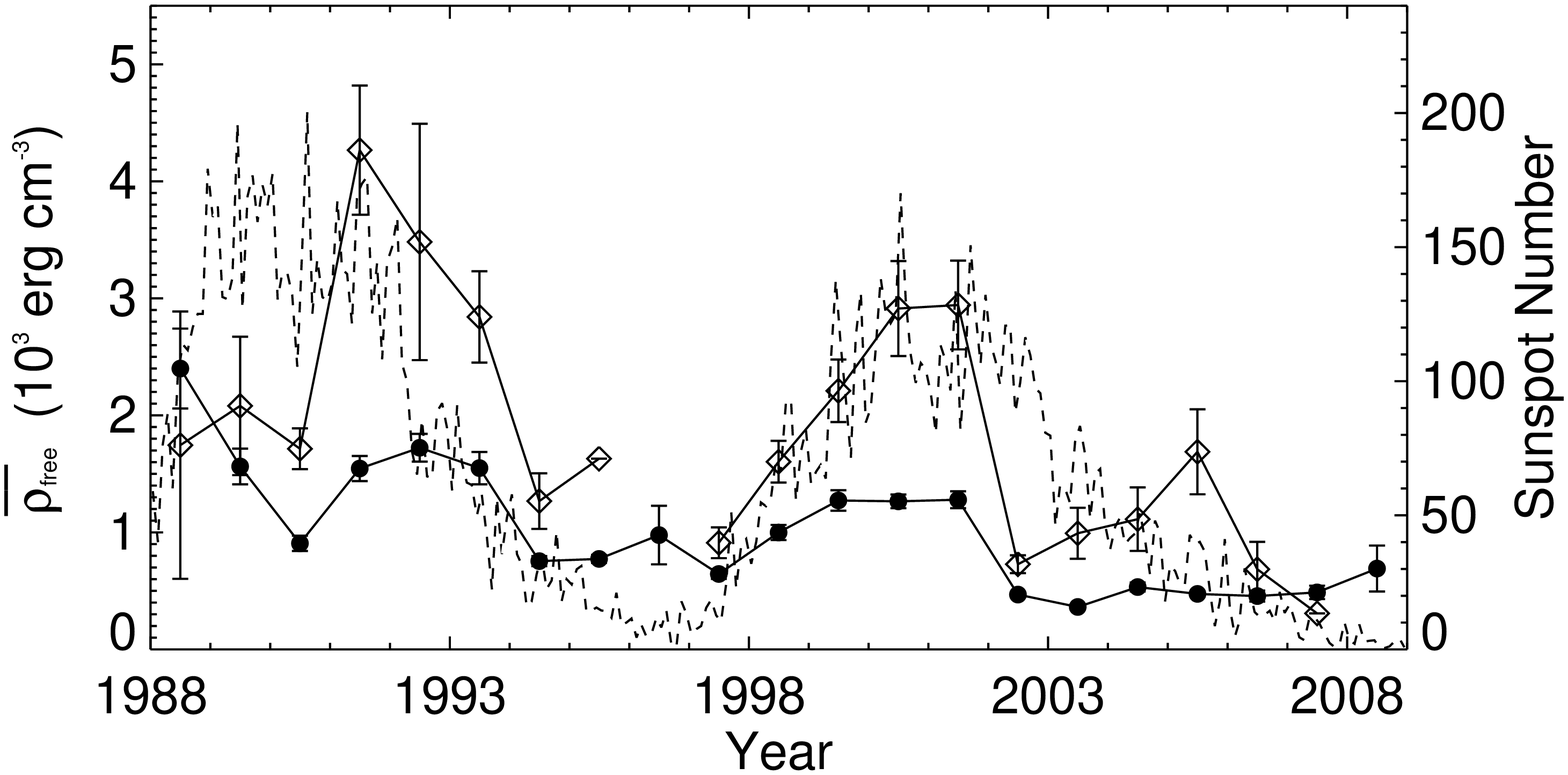}
             }
  \centerline{\hspace*{0.0\textwidth}
              \includegraphics[width=0.525\textwidth,clip=]{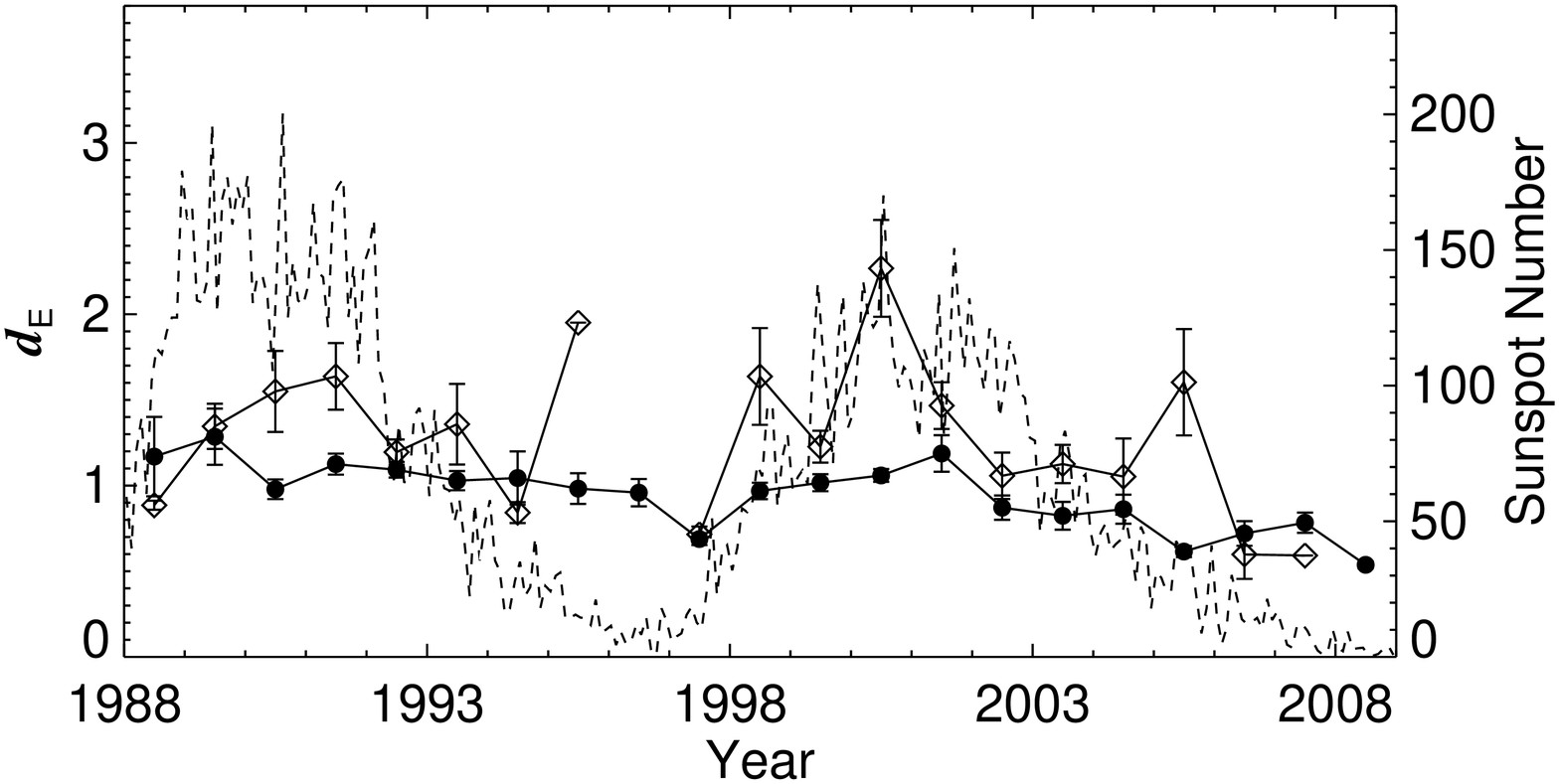}
              \hspace*{-0.025\textwidth}
              \includegraphics[width=0.525\textwidth,clip=]{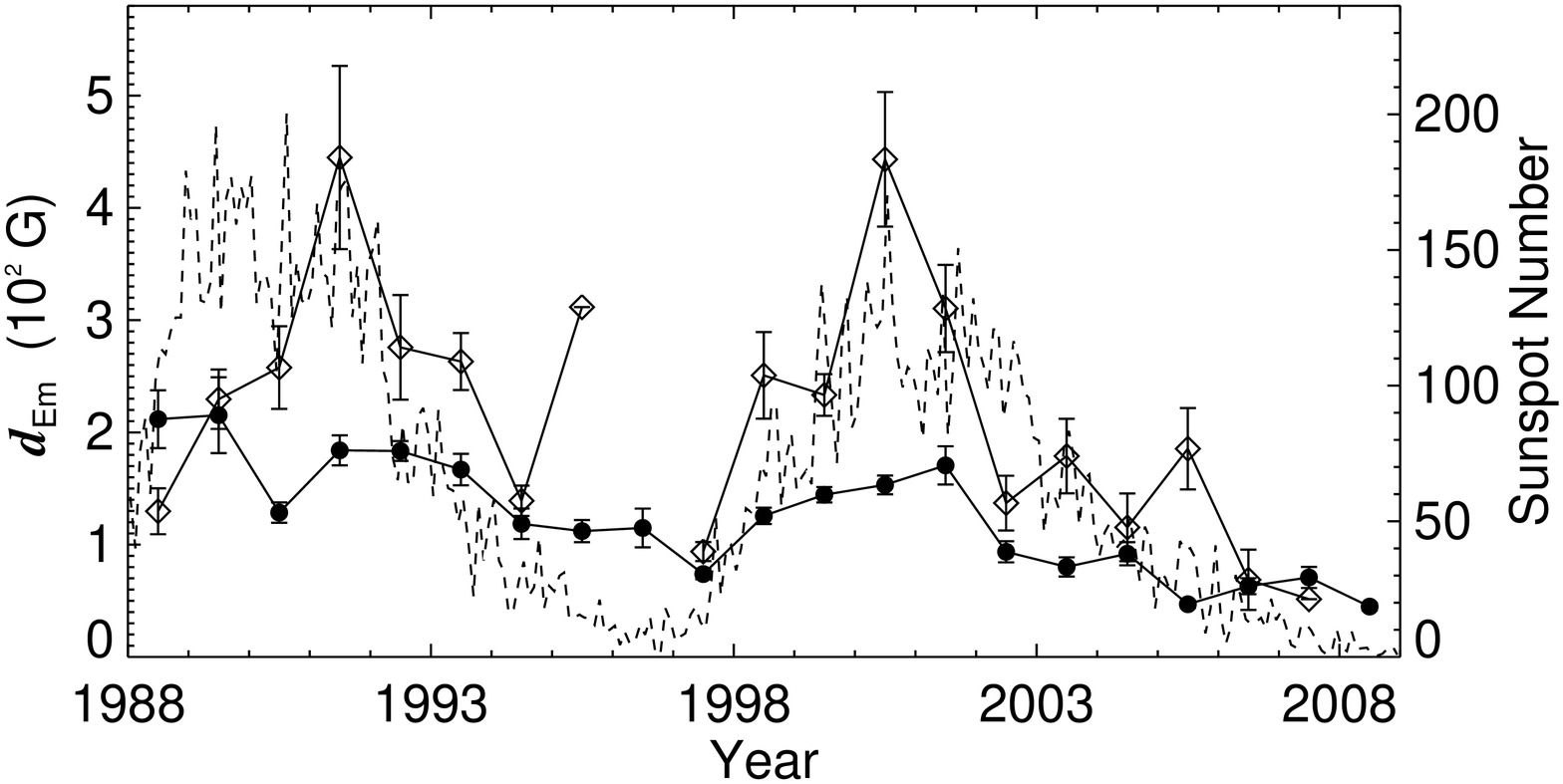}
             }
\caption{Yearly mean values of $\overline{\Delta\phi}$, $\overline{\Delta\psi}$, $\overline{|J_{z}|}$, $\overline{|h_{\rm c}|}$, $|\alpha_{\rm av}|$, $\overline{\rho_{\rm free}}$, $d_{\rm E}$, and $d_{\rm Em}$ of AR samples during 1988--2008. Dots represent the yearly mean values of the samples that did not produce flares with ${\rm FI} \geq 10.0$ in the following 24 h (flare-quiet samples). Diamonds denote the yearly mean values of the samples that produced flares with ${\rm FI} \geq 10.0$ in the following 24 h (flare-productive samples). The monthly mean sunspot numbers during the same period are overlapped (dashed line) for reference.}
     \label{F-meansdev}
\end{figure}

Figure \ref{F-scatpoint} shows the distributions of the above eight parameters of ARs as a function of time from 1988 to 2008. For reference, two plots of the monthly mean sunspot numbers are displayed above other panels. The active samples ({\it i.e.}, flare-productive ARs) are defined as the ARs with the equivalent flare strength ({\it i.e.} flare index FI defined in Section~\ref{S-npwithflare}) greater than a typical value (such as M1.0 here) within the same subsequent time window (24 h in Figure \ref{F-scatpoint}). The rest of them belong to the quiet samples ({\it i.e.}, flare-quiet ARs). The red diamonds denote the active samples, and the black dots represent the quiet samples in Figure \ref{F-scatpoint}. At the same time, the yearly mean values of the active and the quiet samples are calculated separately. Figure \ref{F-meansdev} shows these values. The error bars indicate the standard deviations of the yearly mean values. The plot of monthly mean sunspot numbers (dashed line) is overlaid in each panel in Figure \ref{F-meansdev} for reference.

As presented in Figure \ref{F-scatpoint}, $\overline{\Delta\phi}$ and $\overline{\Delta\psi}$ have similar distributions. The linear correlation coefficient between them is 0.875. Neither of the two shear angles varies with the solar cycle. Most of their values are distributed in the range of $[10\,^\circ,40\,^\circ]$. The points of $\overline{\Delta\psi}$ are more concentrated in this range. The mean value of $\overline{\Delta\psi}$ is $22.0\,^\circ$ with the standard deviation of $6.6\,^\circ$ during 1988--2008, while $\overline{\Delta\phi}$ is $23.4\,^\circ\pm\,7.6\,^\circ$.

Combining with Figure \ref{F-meansdev}, there is no obvious fluctuation of the parameter $\overline{|J_{z}|}$ with the phase of the solar cycle, in which most of AR samples have $\overline{|J_{z}|}$ ranging in 1.0--3.0$\times$10$^{3}$ \textrm{A km$^{-2}$}. Some active samples at solar maximum have larger $\overline{|J_{z}|}$. The samples with $\overline{|J_{z}|}$ larger than 2.5$\times$10$^{3}$ \textrm{A km$^{-2}$} are likely connected with eruptive active regions. During the period from 2001 to 2003, which is the beginning of the declining phase of solar cycle 23, the values of $\overline{|J_{z}|}$ are mostly lower than the general level of 2.0$\times$10$^{3}$ \textrm{A km$^{-2}$}. There are 91.7$\%$ of all the samples in 2001--2003 less than 2.0$\times$10$^{3}$ \textrm{A km$^{-2}$}. After the year 2003, the values of $\overline{|J_{z}|}$ rise to the average level again. Figure \ref{F-meansdev} shows that the distributions of $\overline{|J_{z}|}$ is very similar in form to the two shear angles, $\overline{\Delta\phi}$ and $\overline{\Delta\psi}$.

In the panel of $|\alpha_{\rm av}|$, the values mostly concentrate in the range of 0--1.5$ \times$10$^{-5}$ \textrm{km$^{-1}$}. There are no obvious changes in the parameter $|\alpha_{\rm av}|$ with the the solar cycle. The difference of $|\alpha_{\rm av}|$ between flare-productive and flare-quiet samples is insignificant except in the years 2004 and 2005.

The parameters $\overline{|h_{\rm c}|}$, $\overline{\rho_{\rm free}}$, and $d_{\rm Em}$ follow nicely the variation of solar activity cycle. There are two bell-shaped parts in the corresponding panels during these two solar cycles. The overall level of these parameters rises gradually toward the solar maximum; while near the solar minimum they all fall down. Comparing with the monthly mean sunspot numbers, the troughs of 1989--1990 may be caused by the lack of the available samples or by some unknown uncertainties or evolution processes, but this result is consistent with \inlinecite{Bao98} who have surveyed the evolution of the average current helicity in solar cycle 22. Table \ref{T-lccSN} lists the linear correlation coefficients between these eight parameters and the mean sunspot numbers in the yearly and monthly bases. These coefficients indicate that there is a much closer relationship between the mean sunspot number and $\overline{|h_{\rm c}|}$, $\overline{\rho_{\rm free}}$, and $d_{\rm Em}$.

\begin{table}
\caption{Linear correlation coefficients between eight parameters and the mean sunspot number (SSN), in yearly and monthly bases.}
\label{T-lccSN}
\tabcolsep=3.0pt
\begin{tabular}{ccccccccc}
  \hline
    & $\overline{\Delta\phi}$ & $\overline{\Delta\psi}$ & $\overline{|J_{z}|}$ & $\overline{|h_{\rm c}|}$ & $|\alpha_{\rm av}|$ & $\overline{\rho_{\rm free}}$ & $d_{\rm E}$ & $d_{\rm Em}$ \\
  \hline
Mean SSN (yearly) & 0.379 & 0.010 & 0.093 & 0.594 & $-$0.176 & 0.666 & 0.799 & 0.814 \\
Mean SSN (monthly) & 0.242 & 0.085 & 0.054 & 0.448 & $-$0.067 & 0.474 & 0.367 & 0.530 \\
  \hline
\end{tabular}
\end{table}

Comparing the two parameters $d_{\rm E}$ and $d_{\rm Em}$ representing complexity in magnetic configuration, $d_{\rm E}$ of some ARs in the activity maximum periods is higher than that in the minimum periods, but the number of those ARs with higher $d_{\rm E}$ are not many in the samples, and most of the samples are distributed in a narrow band of $\pm\,0.93$ around $d_{\rm E}=1.04$. Combined with the average strength of magnetic fields, $d_{\rm Em}$ reflects the magnetic complexity of an AR more accurately, while $d_{\rm E}$ is only related to the morphology but limited physics. Both $d_{\rm E}$ and $d_{\rm Em}$ change with the solar cycle, suggesting that there are more complex ARs in the activity maximum periods and less complex ARs in the activity minimum periods.

Using magnetic synoptic charts from Michelson Doppler Imager onboard {\it SOlar and Heliospheric Observatory} (SOHO/MDI) and from National Solar Observatory (NSO/Kitt Peak), \inlinecite{Guo10} found that the proportion of complex ARs (with $d_{\rm E}>1$) to all ARs decreases, although with a large fluctuation, in the declining phase of solar cycle 23, and the figure therein shows a small peak in the year 2005. Investigating the long-term evolution of magnetic helicity in solar cycle 23, \inlinecite{YangSB11} found that the accumulation of magnetic helicity flux has the same trend in both hemispheres around 2005. \inlinecite{ZhangHQ10} and also \inlinecite{Tiwari09} found that ARs in the declining phases of solar cycles 22 and 23 do not follow the general hemispheric helicity sign rule. A similar result has been indicated by \inlinecite{Hao11}. Figures \ref{F-scatpoint} and \ref{F-meansdev} also show a small peak in 2005 in each of the charts. Checking the monthly mean sunspot number and the monthly 10.7cm solar radio flux, there is no abnormal variation around 2005. There are only a little more flares in 2005 and in the end of 2004. Table \ref{T-reldev} shows a comparison between the declining phases of solar cycles 22 and 23. We choose two years with roughly the same mean number of sunspots to compare the corresponding parameters, which means the solar activity level is similar in the two years of each pair. The exception is the mean sunspot number of 2002 and 2003 to compare with the number of 1992. The quantities in Table \ref{T-reldev} represent the relative differences between the values of cycle 23 and cycle 22 ($\frac{x_{23}-x_{22}}{x_{22}}$, where $x_{23}$ and $x_{22}$ represent the yearly mean values of the same parameter in the year of cycle 23 and cycle 22, respectively). The total activity level in the declining phase of cycle 23 is generally lower than that in cycle 22, and coincidentally the parameters in Table \ref{T-reldev} follow the same tendency, that is, the parameter values of cycle 23 are lower than of cycle 22. However, all the parameters of the active samples have larger positive differences between 2005 and 1994. This shows that the nonpotentiality in flare-productive ARs is strong in 2005. A non-monotonic decline of cycle 23 might be regarded as one of the precursors to the long and deep minimum between solar cycles 23 and 24.

\begin{table}
\caption{Comparison of yearly-averaged parameters between the declining phases of solar cycles 22 (1992--1996) and 23 (2002--2007). The tabulated values are relative differences between the values of cycle 23 and cycle 22. AS stands for active samples, QS for quite samples, and TS for total samples. Each of the paired years has comparable yearly mean sunspot numbers.}
\label{T-reldev}
\tabcolsep=2.5pt
\begin{tabular}{crccc||crccc}
  \hline
  & Year & AS & QS & TS &   & Year & AS & QS & TS \\
  \hline
   & 02$\&$03 vs. 92                & $-$0.057 & $-$0.398 & $-$0.329 &   & 02$\&$03 vs. 92                & ~~0.009 & $-$0.269 & $-$0.215 \\
   & 04 vs. 93                      & ~~0.082 & $-$0.335 & $-$0.298 &   & 04 vs. 93                      & ~~0.244 & $-$0.259 & $-$0.214 \\
$\overline{\Delta\phi}$ & 05 vs. 94 & ~~0.407 & $-$0.183 & $-$0.101 &$\overline{\Delta\psi}$ & 05 vs. 94 & ~~0.747 & ~~0.218 & ~~0.291 \\
   & 06 vs. 95                      & ~~0.165 & $-$0.170 & $-$0.126 &   & 06 vs. 95                      & ~~0.372 & ~~0.068 & ~~0.108 \\
   & 07 vs. 96                      & ~~0.295 & ~~0.031 & ~~0.070 &   & 07 vs. 96                      & ~~0.450 & ~~0.129 & ~~0.177 \\
  \hline
   & 02$\&$03 vs. 92             & $-$0.358 & $-$0.560 & $-$0.509 &   & 02$\&$03 vs. 92             & $-$0.603 & $-$0.906 & $-$0.753 \\
   & 04 vs. 93                   & ~~0.092 & $-$0.272 & $-$0.240 &   & 04 vs. 93                   & $-$0.505 & $-$0.609 & $-$0.600 \\
$\overline{|J_{z}|}$ & 05 vs. 94 & ~~0.591 & ~~0.136 & ~~0.199 &$\overline{|h_{\rm c}|}$ & 05 vs. 94 & ~~0.074 & $-$0.442 & $-$0.370 \\
   & 06 vs. 95                   & ~~0.212 & ~~0.086 & ~~0.102 &   & 06 vs. 95                   & $-$0.208 & $-$0.336 & $-$0.320 \\
   & 07 vs. 96                   & $-$0.032 & $-$0.244 & $-$0.212 &   & 07 vs. 96                   & $-$0.493 & $-$0.538 & $-$0.531 \\
  \hline
   & 02$\&$03 vs. 92        & ~~0.891 & $-$0.135 & ~~0.032 &   & 02$\&$03 vs. 92                 & $-$0.609 & $-$1.001 & $-$0.768 \\
   & 04 vs. 93              & ~~1.576 & $-$0.236 & $-$0.074 &   & 04 vs. 93                       & $-$0.415 & $-$0.696 & $-$0.671 \\
$|\alpha_{\rm av}|$ & 05 vs. 94 & ~~6.398 & ~~1.325 & ~~2.028 &$\overline{\rho_{\rm free}}$ & 05 vs. 94 & ~~0.687 & $-$0.407 & $-$0.256 \\
   & 06 vs. 95              & ~~1.980 & ~~0.280 & ~~0.502 &   & 06 vs. 95                       & ~~0.027 & $-$0.467 & $-$0.402 \\
   & 07 vs. 96              & ~~0.192 & $-$0.042 & $-$0.008 &   & 07 vs. 96                       & $-$0.565 & $-$0.502 & $-$0.511 \\
  \hline
   & 02$\&$03 vs. 92 & ~~0.165 & $-$0.282 & $-$0.212 &   & 02$\&$03 vs. 92 & $-$0.105 & $-$0.661 & $-$0.521 \\
   & 04 vs. 93       & $-$0.099 & $-$0.197 & $-$0.189 &   & 04 vs. 93       & $-$0.425 & $-$0.491 & $-$0.485 \\
$d_{\rm E}$ & 05 vs. 94  & ~~0.454 & $-$0.444 & $-$0.320 &$d_{\rm Em}$ & 05 vs. 94 & ~~0.319 & $-$0.637 & $-$0.505 \\
   & 06 vs. 95       & $-$0.350 & $-$0.278 & $-$0.288 &   & 06 vs. 95       & $-$0.432 & $-$0.455 & $-$0.452 \\
   & 07 vs. 96       & $-$0.267 & $-$0.180 & $-$0.193 &   & 07 vs. 96       & $-$0.441 & $-$0.384 & $-$0.393 \\
  \hline
\end{tabular}
\end{table}

\subsection{Nonpotentiality Associated with Flares}
     \label{S-npwithflare}

One of the generally accepted flare classifications is the SXR classification. From 1975 to the present time, the GOES satellite has been recording the whole-Sun X-ray fluxes at 0.5--4 \textrm{\AA} (hard channel) and 1--8 \textrm{\AA} (soft channel) wavelength bands. The SXR flare classification (B, C, M, and X classes) utilizes the flux in the 1--8 \textrm{\AA} range based on the order of magnitude of the peak burst intensity. For analyzing the relationship between the nonpotentiality and solar flares, a time window $\tau$ has been defined as the period forward in time (toward later times) from the observed vector magnetogram. Within a fixed time window $\tau$, the SXR flare index summed by weighting the flares of different classes is used to measure the flaring capability of an AR:
\begin{eqnarray}
{\rm FI}=100\sum_{\tau} I_{\rm X}+10\sum_{\tau} I_{\rm M}+\sum_{\tau} I_{\rm C},
\end{eqnarray}
where $I_{\rm X}$, $I_{\rm M}$, and $I_{\rm C}$ represent the indexes of X-, M-, and C-class SXR flare events, respectively \cite{Antalova96,Abramenko05}. Because the X-ray background in the solar maximum is too high to detect B-class flares ({\it cf.}, \opencite{Feldman97}; \opencite{Joshi10}), the records of B-class SXR flares are not included in the computation. ${\rm FI}=10.0$ (equivalent to a M1.0 flare) is adopted to divide flare-productive ARs from flare-quiet ARs in Section \ref{S-npdistri} and also in the following parts.

In Figure \ref{F-meansdev}, the yearly mean values of each parameter for flare-productive ARs are higher than those for flare-quiet ones in the solar maximum periods, especially in the panels of $\overline{|h_{\rm c}|}$, $\overline{\rho_{\rm free}}$, and $d_{\rm Em}$. However, the twist factor $|\alpha_{\rm av}|$ shows an insignificant difference between the active samples and quiet samples, but shows more apparent difference with much higher uncertainty in 2005. For the same $\tau$, the yearly mean values of each parameter for the flare-productive ARs increase as the threshold of FI is increased, and the differences between flare-productive ARs and flare-quiet ARs are gradually enlarged accordingly. In reverse, for the same threshold of FI, the yearly mean values of each parameter for the flare-productive ARs decrease as the time window $\tau$ is increased, and the differences between flare-productive and flare-quiet ARs are correspondingly a little reduced. Setting a series of $\tau$ and different threshold of FI, the yearly-mean values (Figure \ref{F-meansdev}) will show basically similar distributions and evolution trends for each parameter. However, when the threshold of FI for active samples is raised to a certain limit ({\it e.g.}, M3.0 within 24 h; M5.0 within 48 h), the flare-productive samples vanish in some years of solar minimum periods, especially in the years 1994--1996 and 2006--2008. It suggests that the flare-productive ARs are more likely to have relatively strong nonpotentiality and great complexity, and those parameters characterizing nonpotentiality may be applied as indicators in the flare prediction.

Furthermore, to study the flaring probability of an AR with certain properties of nonpotentiality and complexity, the solar flare productivity is adopted (\opencite{Cui06}, \citeyear{Cui07}; \opencite{Cui08}; \opencite{Park10}), which is defined as
\begin{eqnarray}
P(X)=\frac{N_{\rm A}(\geq{X})}{N_{\rm T}(\geq{X})},
\end{eqnarray}
where $N_{\rm T}(\geq{X})$ is the total number of samples with the values of each parameter greater than its threshold $X$, and $N_{\rm A}(\geq{X})$ is the number of active samples with the values exceeding the same $X$.

\begin{figure}
  \centerline{\hspace*{0.015\textwidth}
              \includegraphics[width=0.425\textwidth,clip=]{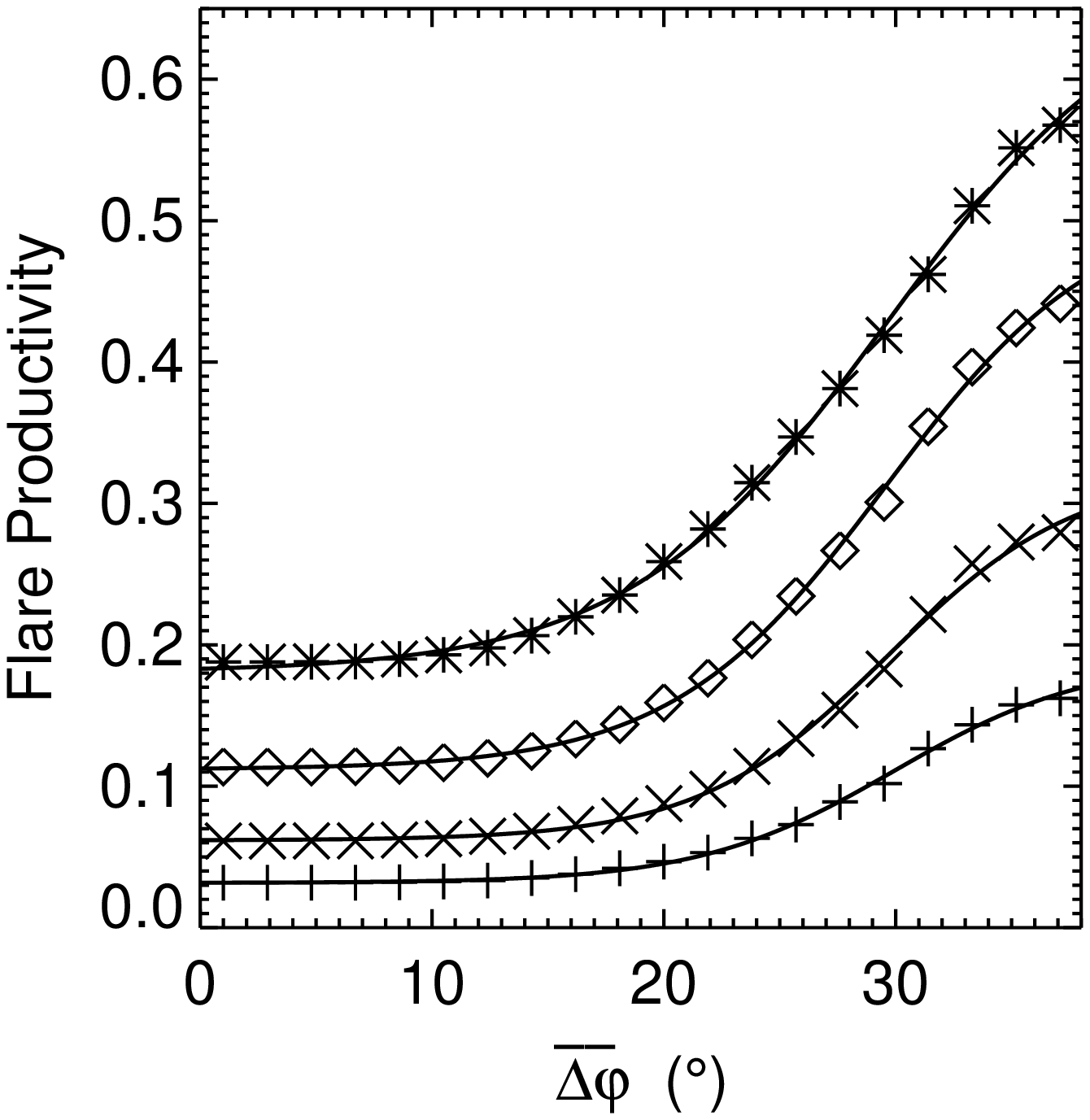}
              \hspace*{-0.03\textwidth}
              \includegraphics[width=0.425\textwidth,clip=]{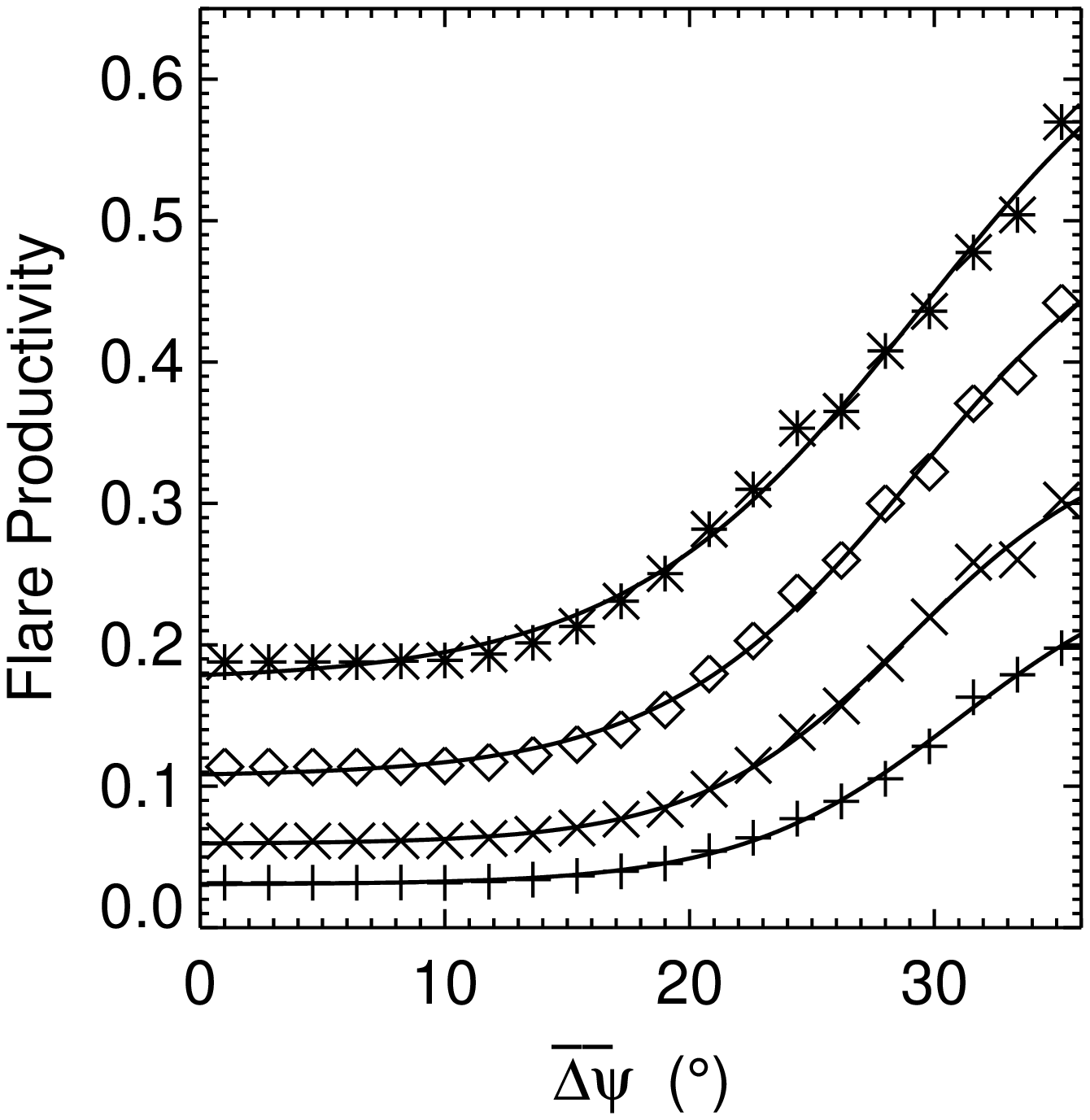}
             }
  \centerline{\hspace*{0.015\textwidth}
              \includegraphics[width=0.425\textwidth,clip=]{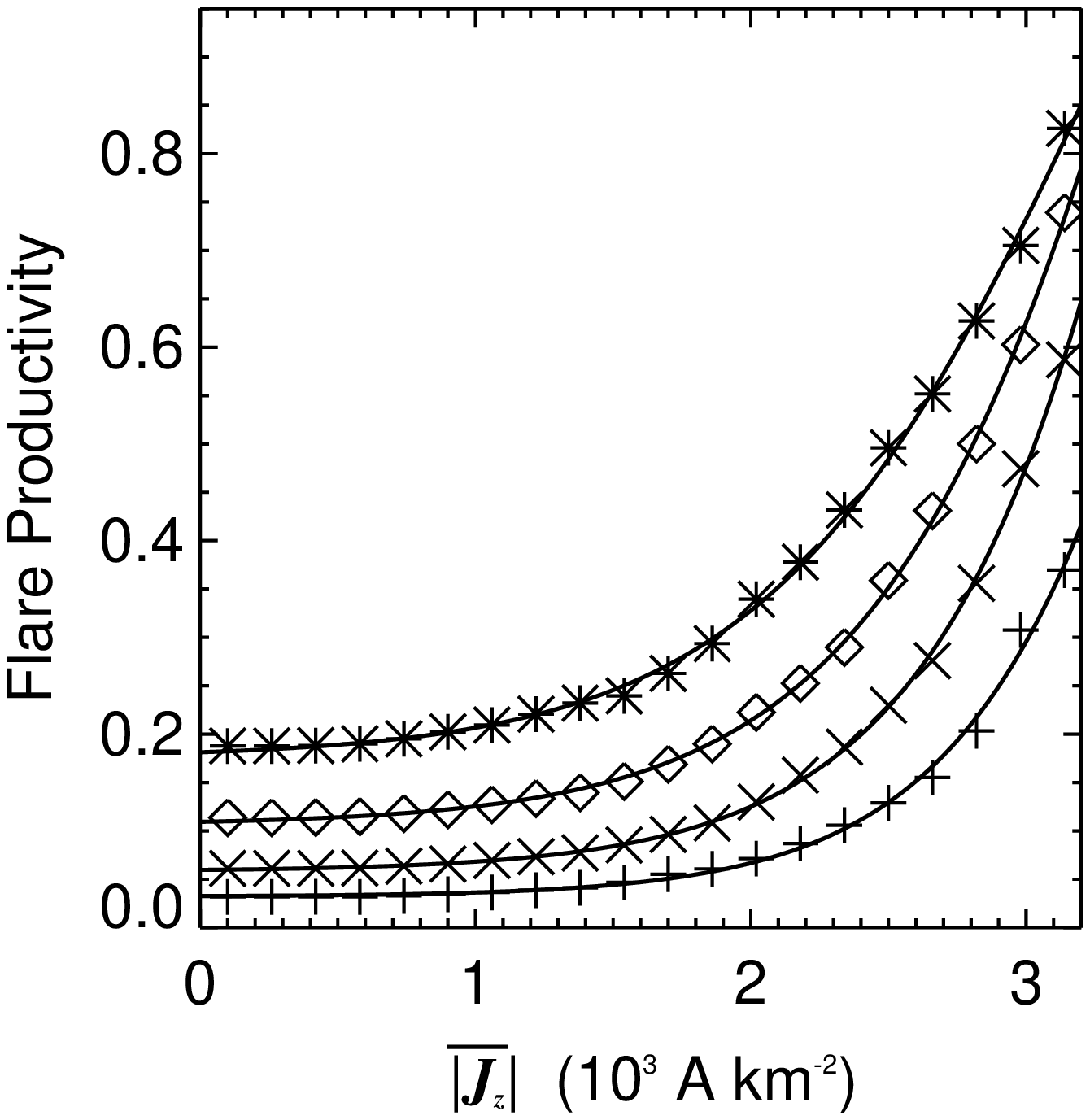}
              \hspace*{-0.03\textwidth}
              \includegraphics[width=0.425\textwidth,clip=]{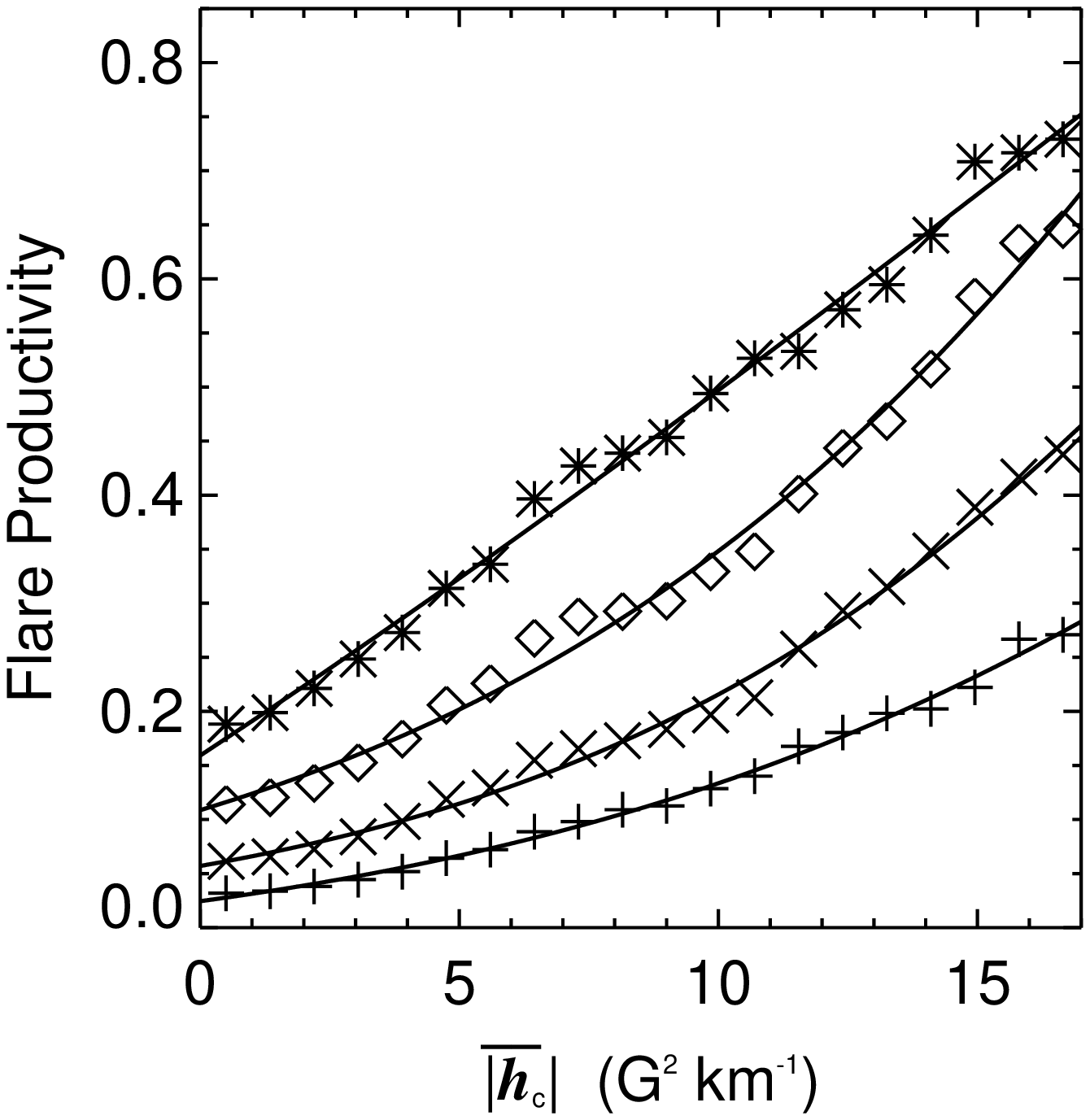}
             }
  \centerline{\hspace*{0.015\textwidth}
              \includegraphics[width=0.425\textwidth,clip=]{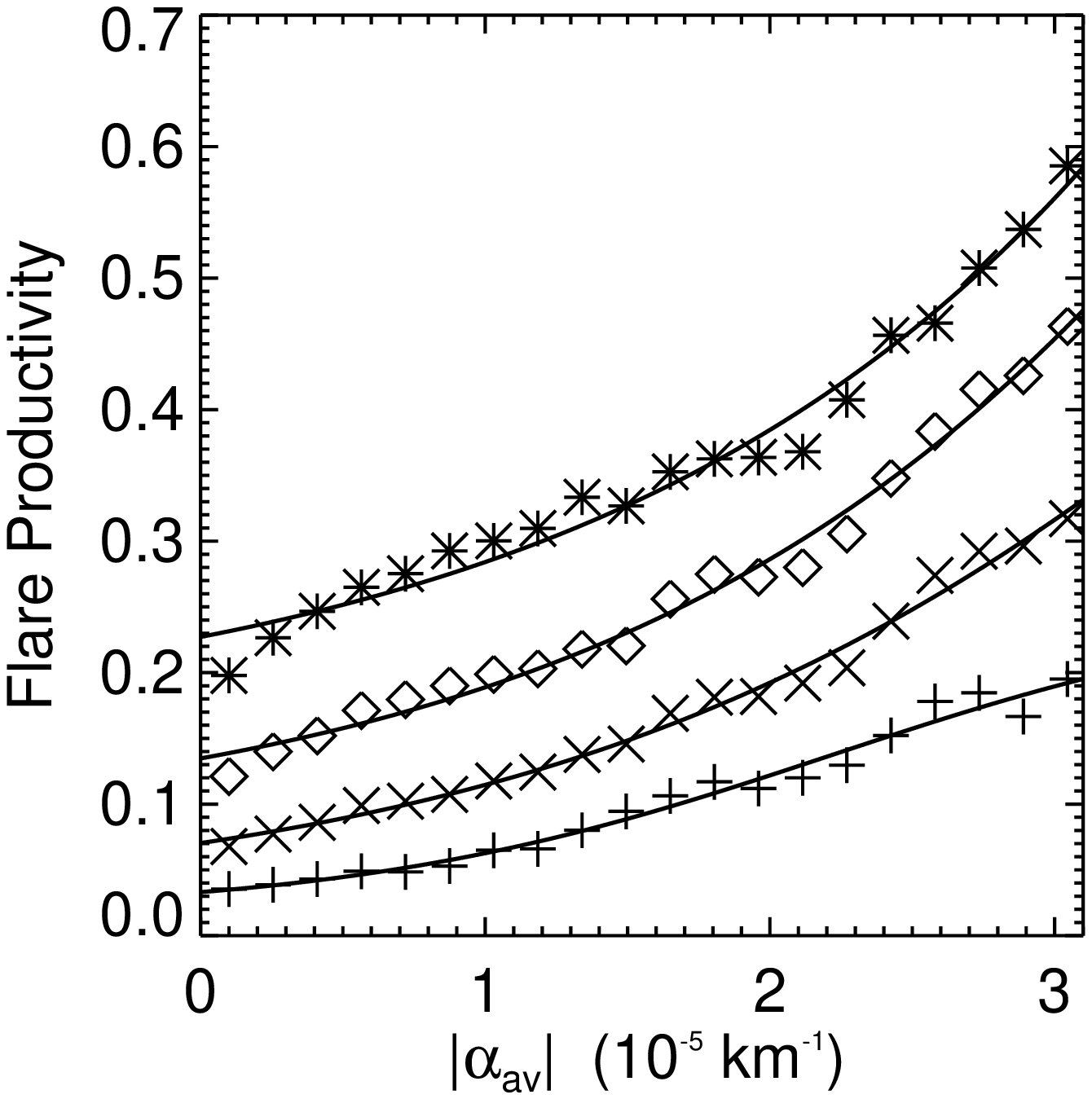}
              \hspace*{-0.03\textwidth}
              \includegraphics[width=0.425\textwidth,clip=]{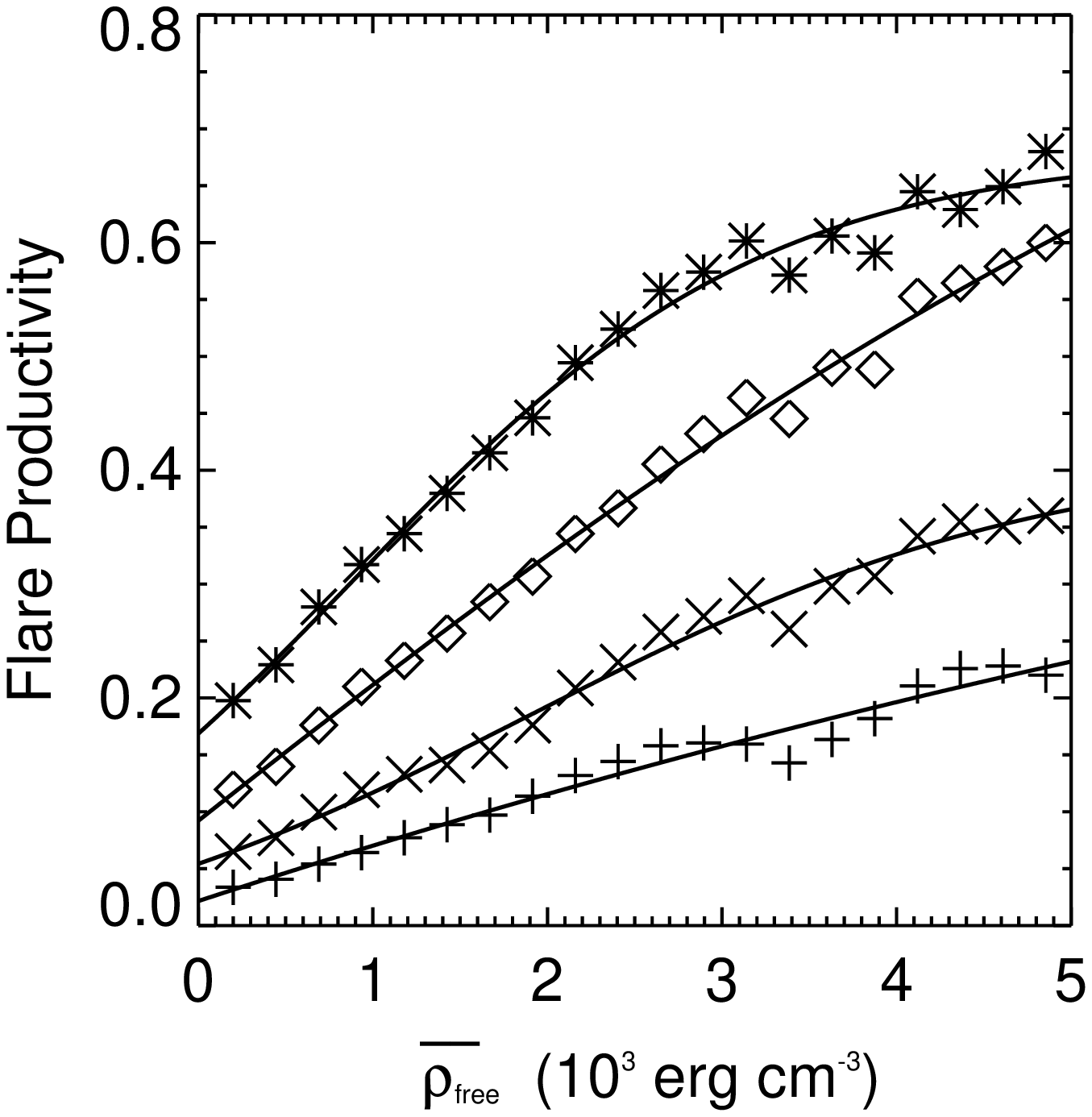}
             }
  \centerline{\hspace*{0.015\textwidth}
              \includegraphics[width=0.425\textwidth,clip=]{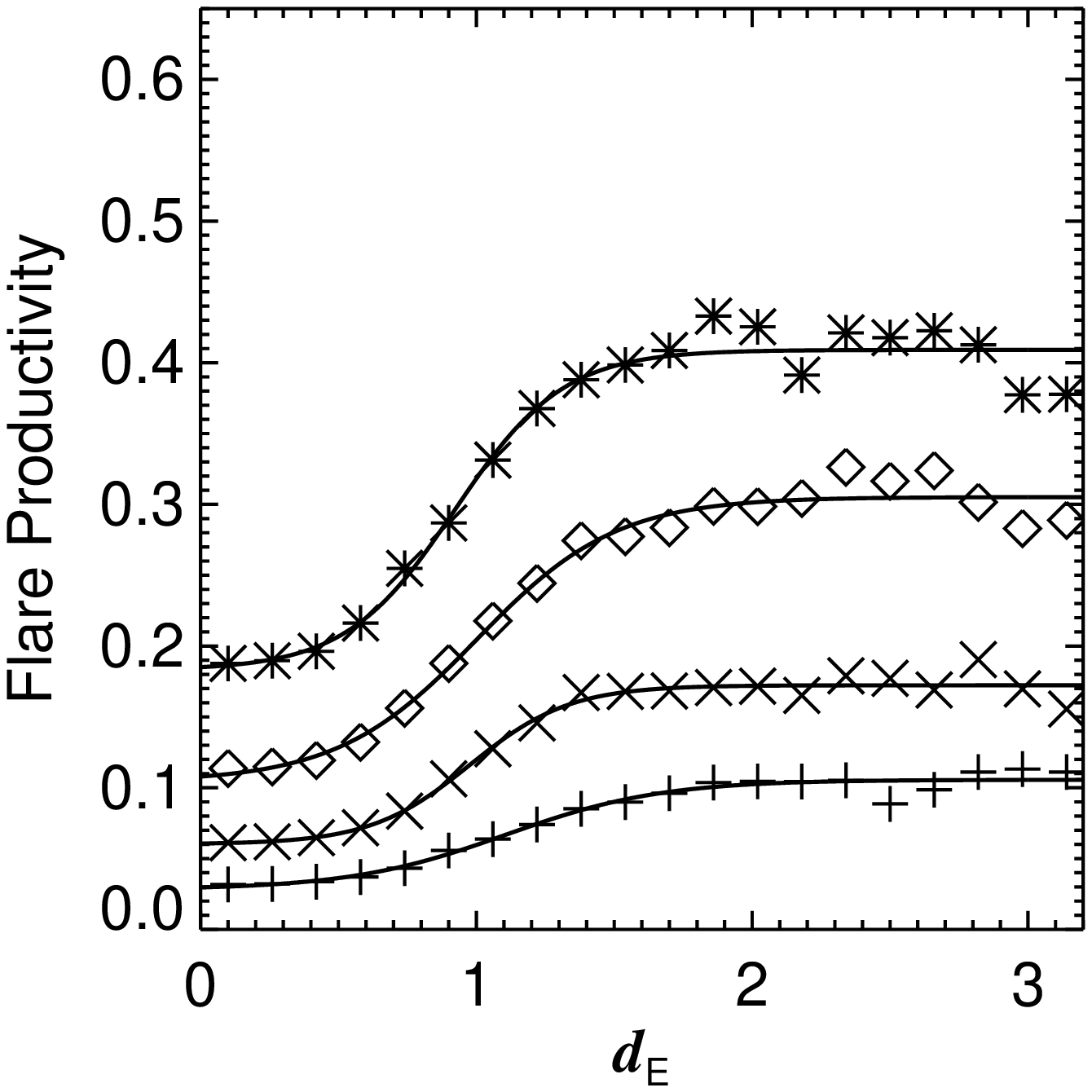}
              \hspace*{-0.03\textwidth}
              \includegraphics[width=0.425\textwidth,clip=]{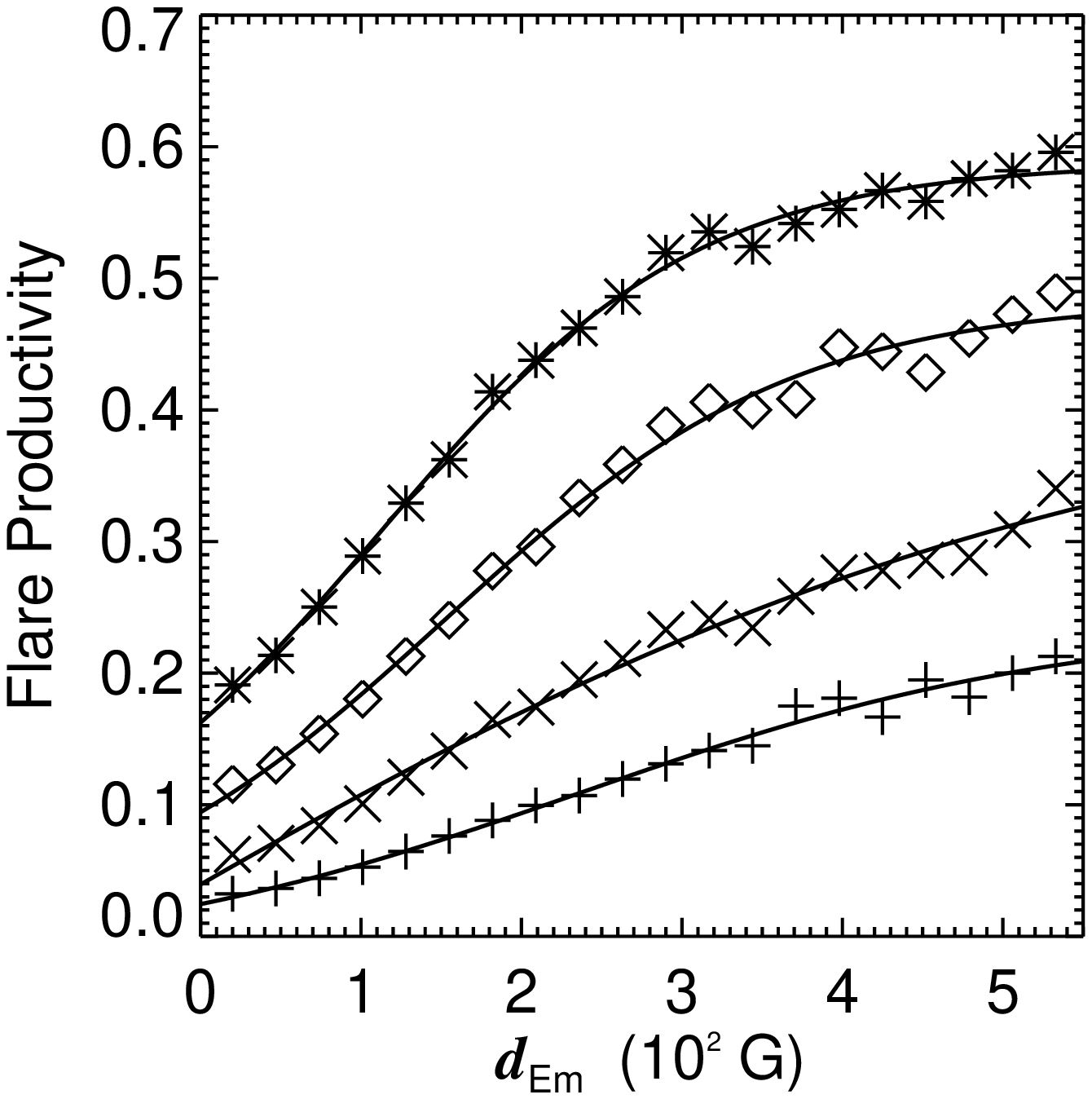}
             }
\caption{Flare productivity for the parameters $\overline{\Delta\phi}$, $\overline{\Delta\psi}$, $\overline{|J_{z}|}$, $\overline{|h_{\rm c}|}$, $|\alpha_{\rm av}|$, $\overline{\rho_{\rm free}}$, $d_{\rm E}$, and $d_{\rm Em}$. The plus, cross, diamond, and asterisk symbols denote the forward time window of 6, 12, 24, and 48 h from the time of the magnetogram, respectively. The solid lines are Boltzmann sigmoid curves fitted to the data.}
     \label{F-flareprod}
\end{figure}

The flare productivities are plotted for each parameter in Figure \ref{F-flareprod}. To guarantee the reliability of the statistics, the total number of samples is no less than 50. In Figure \ref{F-flareprod}, the flare productivities are indicated by plus, cross, diamond, and asterisk symbols when the time window is set as 6, 12, 24, and 48 h from the time of the magnetogram. The Boltzmann sigmoid function is used for fitting the data. $\overline{\Delta\phi}$ and $\overline{\Delta\psi}$ have nearly the same tendency. $\overline{|J_{z}|}$ is also similar to the above two, but rises faster at the value around 2.5$\times$10$^{3}$ \textrm{A km$^{-2}$}. $\overline{|h_{\rm c}|}$ and $|\alpha_{\rm av}|$ have similar increasing tendency in this plot. The shape of the fitting curve of $\overline{\rho_{\rm free}}$ bears a resemblance to that of $d_{\rm Em}$. When the value of $d_{\rm E}$ is greater than 1.0, there are almost no distinct differences in the flaring capability. These plots reveal different flare-productivity levels and their respective variation tendencies, as the result of the distributions and variations of the eight parameters in Figures \ref{F-scatpoint} and \ref{F-meansdev}. In Figure \ref{F-flareprod}, the fitting curves of the parameters with less differences between flare-productive ARs and flare-quiet ARs are relatively flatter than those with more differences. The largest flare productivity versus the largest $X$ values in all the panels of Figure \ref{F-flareprod} may reflect their respective contributions to the flare prediction, and these specific parameters could be given corresponding weight in prediction models.

\subsection{Comparison among Nonpotentiality Parameters and Effective Distances}
     \label{S-compareParam}

\begin{table}
\caption{Linear correlation coefficients for eight nonpotentiality and complexity parameters.}
     \label{T-lcc}
\tabcolsep=5.0pt
\begin{tabular}{ccccccccc}
  \hline
    & $\overline{\Delta\phi}$ & $\overline{\Delta\psi}$ & $\overline{|J_{z}|}$ & $\overline{|h_{\rm c}|}$ & $|\alpha_{\rm av}|$ & $\overline{\rho_{\rm free}}$ & $d_{\rm E}$ & $d_{\rm Em}$ \\
  \hline
$\overline{\Delta\phi}$ & 1.000 & -- & -- & -- & -- & -- & -- & -- \\
$\overline{\Delta\psi}$ & 0.875 & 1.000 & -- & -- & -- & -- & -- & -- \\
$\overline{|J_{z}|}$ & 0.528 & 0.499 & 1.000 & -- & -- & -- & -- & -- \\
$\overline{|h_{\rm c}|}$ & 0.386 & 0.274 & 0.737 & 1.000 & -- & -- & -- & -- \\
$|\alpha_{\rm av}|$ & 0.306 & 0.425 & 0.285 & 0.107 & 1.000 & -- & -- & -- \\
$\overline{\rho_{\rm free}}$ & 0.372 & 0.313 & 0.566 & 0.911 & 0.101 & 1.000 & -- & -- \\
$d_{\rm E}$ & 0.259 & 0.186 & 0.162 & 0.220 & 0.076 & 0.242 & 1.000 & -- \\
$d_{\rm Em}$ & 0.330 & 0.241 & 0.357 & 0.566 & 0.069 & 0.618 & 0.856 & 1.000 \\
  \hline
\end{tabular}
\end{table}

\begin{figure}
  \centerline{\hspace*{0.015\textwidth}
              \includegraphics[width=0.48\textwidth,clip=]{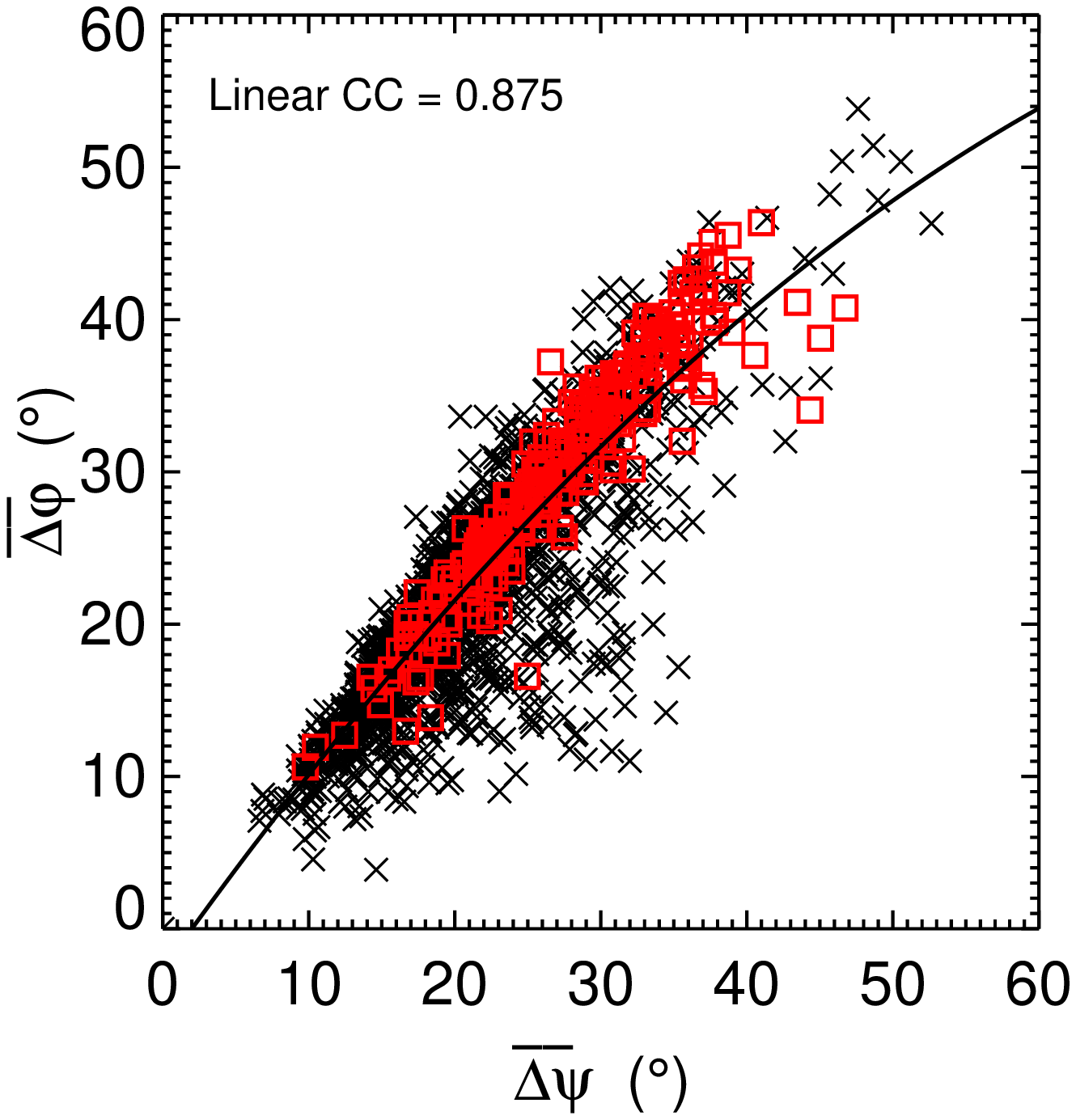}
              \hspace*{-0.045\textwidth}
              \includegraphics[width=0.48\textwidth,clip=]{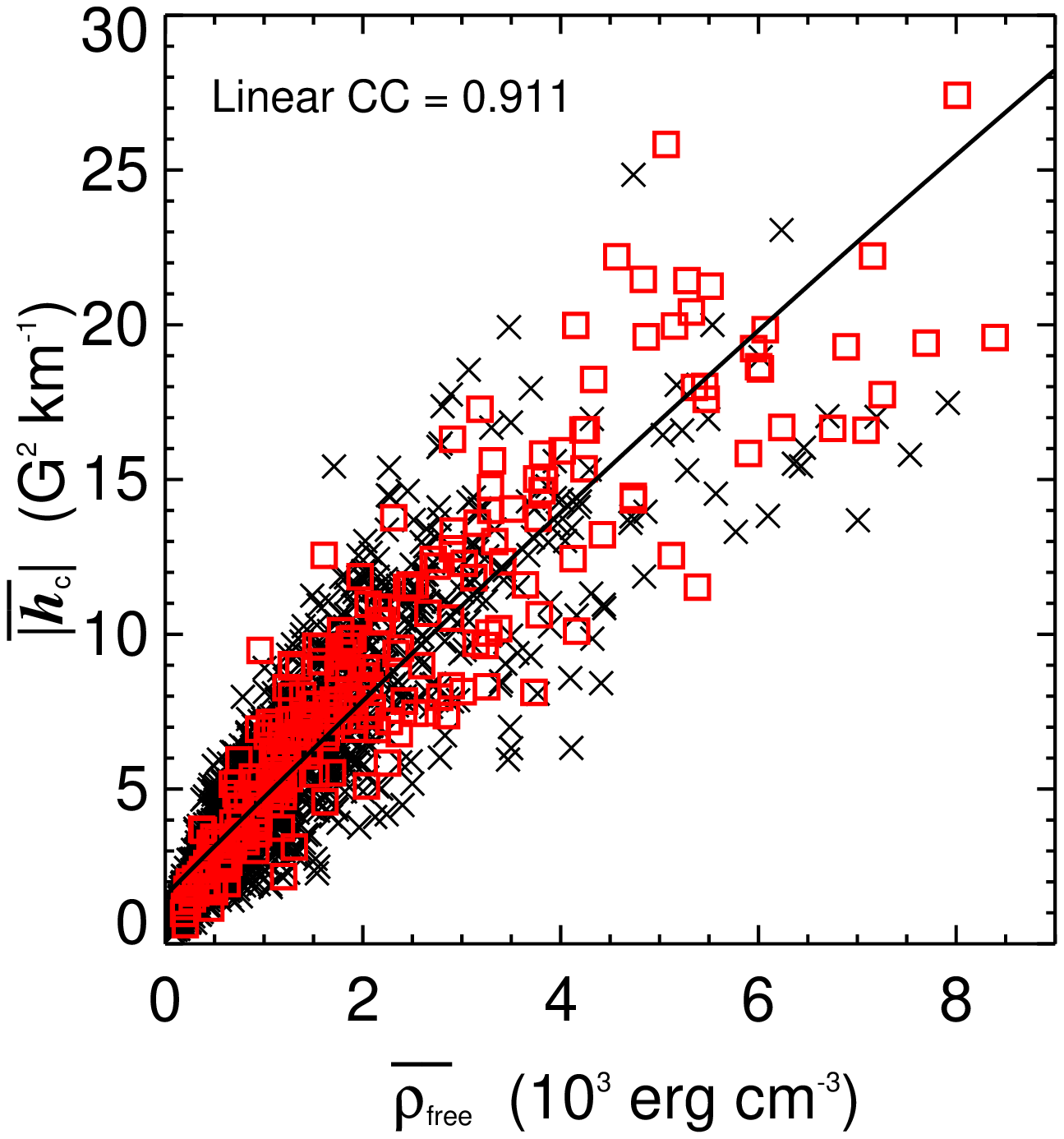}
             }
  \centerline{\hspace*{0.015\textwidth}
              \includegraphics[width=0.48\textwidth,clip=]{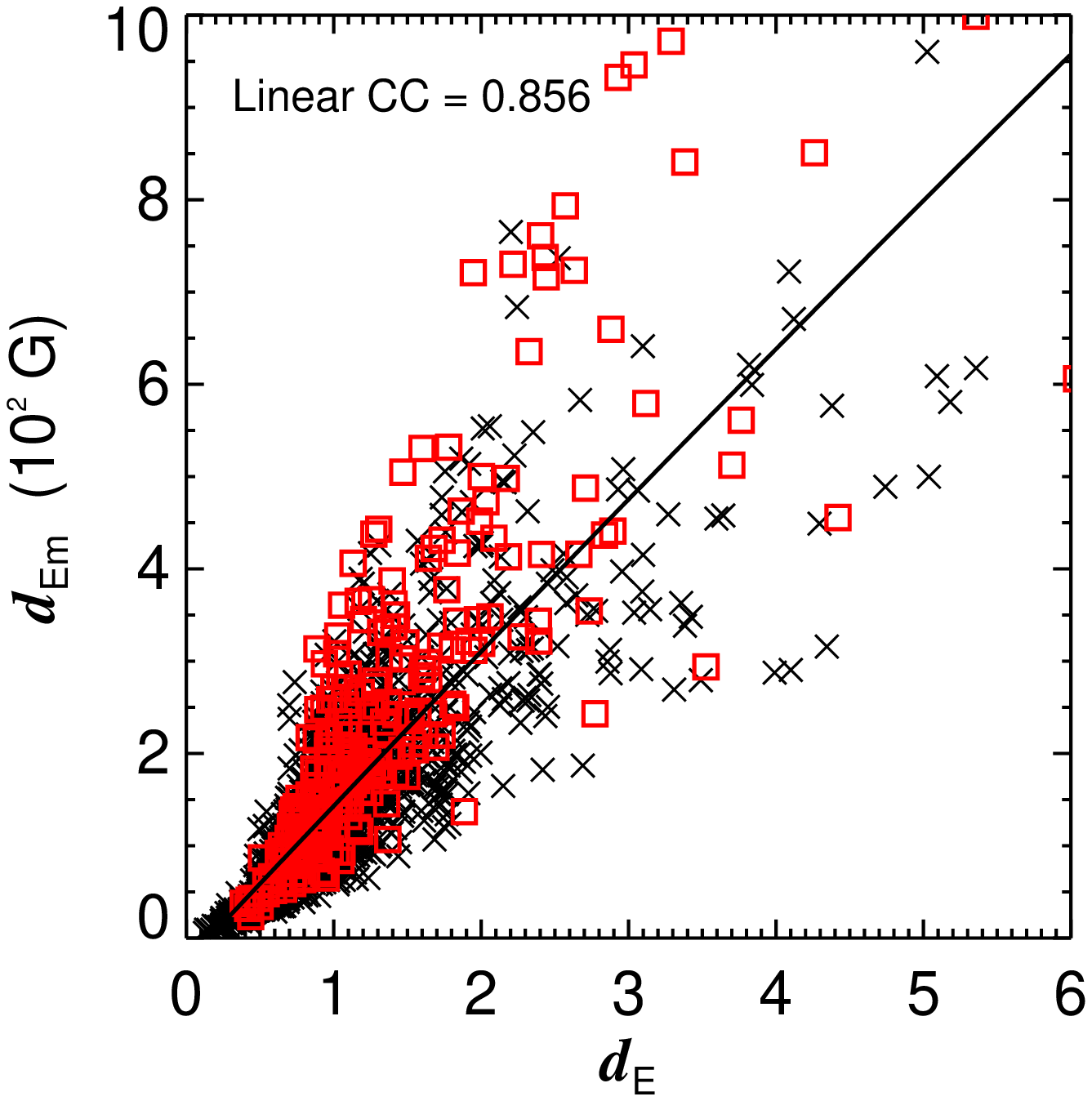}
              \hspace*{-0.045\textwidth}
              \includegraphics[width=0.48\textwidth,clip=]{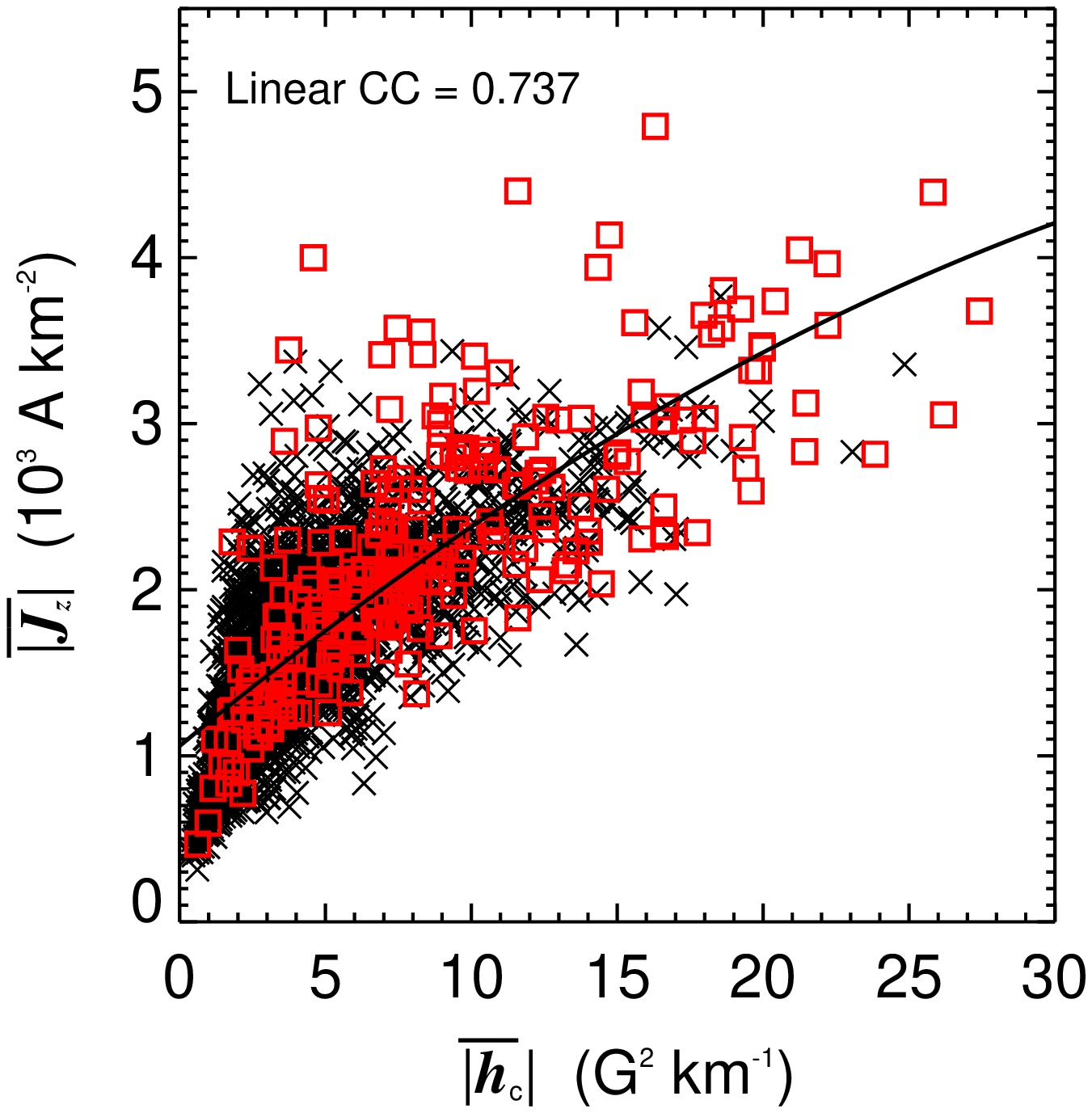}
             }
\caption{Correlations between $\overline{\Delta\phi}$ and $\overline{\Delta\psi}$, $\overline{|h_{\rm c}|}$ and $\overline{\rho_{\rm free}}$, $d_{\rm E}$ and $d_{\rm Em}$, $\overline{|h_{\rm c}|}$ and $\overline{|J_{z}|}$. In each panel, the black crosses denote the flare-quiet samples, and the red squares denote the flare-productive samples; the solid line is the least-square quadratic polynomial fitting curve. `Linear CC' stands for the linear correlation coefficient, and its value is obtained from all the data points of a panel.}
     \label{F-lccpfit}
\end{figure}

The linear correlation coefficients for every two nonpotentiality and complexity parameters are listed in Table \ref{T-lcc}. Being consistent with the results mentioned in Sections \ref{S-npdistri} and \ref{S-npwithflare}, $d_{\rm Em}$ has a relatively close relation with $\overline{|h_{\rm c}|}$ and $\overline{\rho_{\rm free}}$. $d_{\rm Em}$ has relatively higher positive correlation with nonpotentiality than $d_{\rm E}$, while $d_{\rm E}$ have almost no relation with nonpotential parameters. The directly related parameters $\overline{\Delta\phi}$ and $\overline{\Delta\psi}$, $d_{\rm E}$ and $d_{\rm Em}$, and $\overline{|h_{\rm c}|}$ and $\overline{|J_{z}|}$ are highly correlated with each other as expected. The parameters $\overline{|h_{\rm c}|}$ and $\overline{\rho_{\rm free}}$ also show very high correlation. Figure \ref{F-lccpfit} displays scatter plots of these four pairs which show the highest correlation coefficients, 0.911, 0.875, 0.865, and 0.737. The flare-productive samples and flare-quiet samples are plotted as red squares and black crosses, respectively. The least-square quadratic polynomial fitting is made for all the data points of each panel.

\section{Conclusions}
     \label{S-conclusion}

By calculating eight parameters ($\overline{\Delta\phi}$, $\overline{\Delta\psi}$, $\overline{|J_{z}|}$, $\overline{|h_{\rm c}|}$, $|\alpha_{\rm av}|$, $\overline{\rho_{\rm free}}$, $d_{\rm E}$, and $d_{\rm Em}$) of nonpotentiality and complexity for 2173 photospheric vector magnetograms in 1106 ARs associated with flares, we found the main results as follows:

\begin{description}
  \item
(1) On average, two mean magnetic shear angles $\overline{\Delta\phi}$ and $\overline{\Delta\psi}$, mean absolute vertical current density $\overline{|J_{z}|}$, absolute twist factor $|\alpha_{\rm av}|$, and effective distance $d_{\rm E}$ in ARs do not change significantly with the global solar activity level. However, it is more likely that these parameters show higher values in the solar maximum than in the solar minimum.
  \item
(2) The mean absolute current helicity density $\overline{|h_{\rm c}|}$, mean free magnetic energy density $\overline{\rho_{\rm free}}$, and modified effective distance $d_{\rm Em}$ show high positive correlation with the mean sunspot number, and these parameters also have relatively close relationship with each other. The Pearson linear correlation coefficients of the above three with the yearly mean sunspot numbers are larger than 0.59. They can be used to characterize the solar activity level as well as the traditional sunspot number.
  \item
(3) The nonpotentiality and complexity parameters between flare-productive ARs and flare-quiet ones computed in this work are useful to understand the evolution of flare-productive ARs and the relationship with the magnetic activity levels of the cycles. These nonpotentiality and complexity parameters may be synthetically applied as indicators to predict solar flares with some weight.
  \item
(4) Due to the loss of the information of magnetic field strength in the parameter of effective distance $d_{\rm E}$, the modified effective distance $d_{\rm Em}$ (including the strength of the magnetic field) turns out to be much better in indicating the magnetic activities of ARs.
\end{description}

The accumulated nonpotential energy in the magnetic field of solar ARs provides the energy of solar flares, while the trigger of flares is also related to the magnetic nonpotentiality. The parameters discussed above will provide the basic information of nonpotentiality of solar active regions in different aspects. The synoptic analysis of different nonpotential parameters may supply an objective basis on the relationship between the nonpotential field and solar flares in order to predict solar flares.

Although long-term accumulation of the data from ground-based instruments is indispensable, the vector magnetograms taken by the {\it Hinode} satellite, {\it Solar Dynamics Observatory} (SDO), and other space-borne solar telescopes without being affected from atmospheric seeing have been giving more accurate data with better spatial resolution for further comparison studies.

%

%
\begin{acks}
The authors wish to thank the referee and editor for important comments and suggestions. One of the authors X.Yang is grateful to Prof. X.J.Mao and Dr. J.Jiang for reading through the manuscript and giving beneficial advices. This work is supported by the National Natural Science Foundation of China (10733020, 10921303, 11003025, 11103037, 11103038, 60940030, 11173033, and 41174153), the Young Researcher Grant of National Astronomical Observatories of Chinese Academy of Sciences (CAS), the Knowledge Innovation Program of CAS (KJCX2-EW-T07), and the Key Laboratory of Solar Activity of CAS. The authors are grateful to Mrs. G.P.Wang and the HSOS staff for producing the nice data, and also the GOES team for the helpful records.
\end{acks}

%
%
\bibliographystyle{spr-mp-sola-cnd} 
\bibliography{SOLA1731}
%
%
%

\end{article}

\end{document}